\documentclass[10pt,journal]{IEEEtran}
\usepackage{amsmath,amssymb,amsfonts}
\usepackage{algorithmic}
\usepackage{graphicx}
\usepackage{textcomp}
\usepackage{xcolor, comment}
\usepackage{tabularx}
\usepackage{subcaption}
\usepackage{caption}
\usepackage{url}
\usepackage{float}
\usepackage{pifont}
\usepackage{acronym}
\usepackage{balance}
\usepackage{array}

\newcommand{\xmark}{\ding{55}}%
\newcolumntype{P}[1]{>{\centering\arraybackslash}p{#1}}

\acrodef{ML}{Machine Learning}
\acrodef{LPCC}{Linear Predictive Cepstral Coefficients}
\acrodef{MFCC}{Mel Frequency Cepstral Coefficients}
\acrodef{AWGN}{Additive White Gaussian Noise}
\acrodef{SNR}{Signal-to-Noise Ratio}
\acrodef{UAS}{Unmanned Aerial System}
\acrodef{UAV}{Unmanned Aerial Vehicle}
\acrodef{FFT}{Fast Fourier Transform}
\acrodef{SDR}{Software Defined Radio}
\acrodef{TDoA}{Time Difference of Arrival}
\acrodef{SVM}{Support Vector Machine}
\acrodef{STFT}{Short-time Fourier transform}
\acrodef{RPM}{Revolutions Per Minute}
\acrodef{IoT}{Internet of Things}
\acrodef{KNN}{k-Nearest Neighbors}
\acrodef{RF}{Radio-Frequency}
\acrodef{GNSS}{Global Navigation Satellites System}

\def\BibTeX{{\rm B\kern-.05em{\sc i\kern-.025em b}\kern-.08em
    T\kern-.1667em\lower.7ex\hbox{E}\kern-.125emX}}

\title{Noise2Weight:\\ On Detecting Payload Weight from Drones Acoustic Emissions}

\author{
    \IEEEauthorblockN{Omar Adel Ibrahim\IEEEauthorrefmark{1},
                      Savio Sciancalepore\IEEEauthorrefmark{2} and
                      Roberto Di Pietro\IEEEauthorrefmark{3}}
    \IEEEauthorblockA{\\Division of Information and Computing Technology,\\
                      College of Science and Engineering,\\
                      Hamad Bin Khalifa University, Doha, Qatar\\
Email: \IEEEauthorrefmark{1}oaibrahim@mail.hbku.edu.qa,
      \IEEEauthorrefmark{2}ssciancalepore@hbku.edu.qa,
      \IEEEauthorrefmark{3}rdipietro@hbku.edu.qa}
}

\begin{document}
\bstctlcite{IEEEexample:BSTcontrol}

\maketitle

\begin{abstract}

The increasing popularity of autonomous and remotely-piloted drones have paved the way for several use-cases, e.g., merchandise delivery and surveillance.  In many scenarios, estimating with zero-touch the weight of the payload carried by a drone before its physical approach could be attractive, e.g., to provide an early tampering detection.

In this paper, we investigate the possibility to remotely detect the weight of the payload carried by a commercial drone by analyzing its acoustic fingerprint. We characterize the difference in the thrust needed by the drone to carry different payloads, resulting in significant variations of the related acoustic fingerprint. 
We applied the above findings to different use-cases, characterized by different computational capabilities of the detection system. Results are striking: using the Mel-Frequency Cepstral Coefficients (MFCC) components of the audio signal and different Support Vector Machine (SVM) classifiers, we achieved a minimum classification accuracy of $98\%$ in the detection of the specific payload class carried by the drone, using an acquisition time of $0.25$~s---performances improve when using longer time acquisitions. 

All the data used for our analysis have been released as open-source, to enable the community to validate our findings and use such data as a ready-to-use basis for further investigations.

\end{abstract}




\section{Introduction}

The last few years have witnessed the exponential diffusion of the \ac{UAV} technology~\cite{altawy2016security}. \acp{UAV}, also referred to as \emph{drones}, are nowadays pervasive, opening up to advances and innovations in a variety of use-cases and application scenarios, including perimeter control, remote surveillance, emergency management, merchandise delivery, and military applications, to name a few~\cite{ullah2020_access}. Considering only the logistic transportation sector, specialized agencies estimate an increase of the market size of more than 29 billion USD from 2022 to 2027~\cite{dronemarket}, and popular companies such as Amazon are already testing platforms to deliver merchandise to selected customers using drones~\cite{amazonAir}.

Unfortunately, drones represent the classical dual-use technology that can also be used by malicious entities. Indeed,
several attacks have been already realized using remotely-piloted or autonomous drones, equipped with explosive charges and weapons~\cite{bbcDrone},~\cite{hauthi},~\cite{vatta2016_icwmc}. At the same time, leading merchandise industries are delaying the introduction of drones, due to increasing concerns about the safety of the surrounding people, as well as tampering and theft of both the drone and the carried merchandise~\cite{forbes_dronedelivery}.

The above concerns motivated several contributions from the scientific community. On the one hand, a variety of solutions for the detection of an approaching drone have been introduced, based either on radar~\cite{hoffmann2016micro}, visual target recognition~\cite{Nassi_2019}, \ac{RF} analysis~\cite{Nguyen2017}, traffic analysis~\cite{sciancalepore2020pinch}, and sound analysis~\cite{anwar2019machine}. 
On the other hand, a few solutions to secure drone merchandise operations have been proposed, mainly focused on the detection of the most common RF attacks, such as jamming and spoofing of \ac{GNSS} technologies~\cite{tedeschi2020_access},~\cite{oligeri2019_wisec},~\cite{oligeri2020_wisec}.

Concerning the analysis of the sound produced by a drone, the rationale behind its application is that the acoustic noise produced by the motors and the blades of the drones represents a unique feature, clearly distinguishing drones from other targets, as well as drones from other drones. Therefore, audio analysis has been used for the detection of the drone in an outdoor environment~\cite{anwar2019machine}, as well as to discriminate the particular brands and models of drones~\cite{shi2018hidden}.

However, to the best of our knowledge, none of the previous scientific contributions investigated the potential of using acoustic emissions generated by a drone to infer the weight of the carried payload. Such a zero-touch detection method can have several heterogeneous applications. In the merchandising sector, a customer equipped with a dedicated system can realize if the weight of the payload carried by the drone is consistent with the weight of the expected goods. If the detected value differs from the expected one, he could refuse the drone to approach, limiting possible disputes about the delivery.  In the military sector, it could be possible to verify if a drone used for autonomous or remotely-piloted exploration has been captured and tampered with---for instance, equipped with explosive materials. 
All these applications could leverage the previous knowledge of the specific brand and model of the drone, and could take place when the drone is hovering in front of the acquisition system. In addition, when coupled with existing methods for drone brand and model identification, using the acoustic emissions for payload weight detection could increase the capability to detect in advance threats to critical infrastructures.

{\bf Contribution.}  To the best of our knowledge, we are the first to investigate the existing tight correlation between the acoustic noise produced by a commercial drone and the weight of the carried payload. We run an extensive experimental campaign over a reference 3DR Solo drone, and we verified that the adjustments in the speed of the motors and blades of the drone, required to adapt the thrust to different payload weights, leads to a change in the profile of the acoustic emissions of the drone. This side-channel can be used to infer on the weight of the payload carried by the drone, by using different techniques.

We presented two scenarios, characterized by either constraints on the capabilities of the devices used for the analysis of the sound, or by regular processing capabilities, such as the ones of a regular laptop. 
We showed that a system using only the pitch of the sound can discriminate if a drone is carrying a payload or not, with an accuracy that depends on both the size of the acquisition window and the specific payload carried by the drone. Conversely, a system using the \ac{MFCC} representation of the audio signal and standard supervised \ac{ML} techniques can detect the approximate payload carried by the drone (with an accuracy of $50$~g), with a minimum classification accuracy of $98$\%, acquiring the acoustic emissions for only $0.25$~s. Further analyses on the effect of the noise strengthen the validity of our findings. The data used for our analysis have been publicly-released as open-source, to enable interested researchers to verify our claims and to employ these data as a ready-to-use basis for further investigations~\cite{data}.

{\bf Roadmap.} The rest of the paper is organized as follows: Section \ref{sec:rw} reviews the related work, Section \ref{sec:scenario} discusses the different scenarios suggested for this work, Section \ref{sec:setup} introduces the experimental setup used in this work, Section \ref{sec:sounds} provides the details of the components of the sound that are critical for the identification of the payload carried by the drone, while Section \ref{sec:results} includes the results of our investigation in the identified reference scenarios, as well as an analysis of the impact of the noise. Finally, Section \ref{sec:conc} concludes the paper.

\section{Related Work}
\label{sec:rw}

A few contributions in the last years investigated the possibility to identify the presence of a drone via its acoustic emissions. For instance, the authors in \cite{anwar2019machine} designed an \ac{ML} framework that utilizes features extracted using the linear \ac{LPCC} and \ac{MFCC} techniques to detect and classify amateur drone sounds, compared to a few other sound sources. The authors collected a small dataset consisting of 172 Samples (maximum duration of 55s) across four categories of sounds (i.e., drones, thunderstorms, birds and airplanes) in a noisy environment, using an array of microphones. Overall, they achieved 96.7\% detection accuracy using \ac{MFCC} only (as it outperforms \ac{LPCC}) and the cubic SVM classification algorithm.

The authors in  \cite{bernardini2017drone} used a combination of time and frequency domain features to fingerprint the drones' sounds. They collected around 850 samples of each of some different environmental sounds such as nature daytime, train passing, crowd, street with traffic, in addition to the drone sounds. Then, the features extracted from these recordings were passed to an \ac{SVM} algorithm using the Gaussian Radial Basis kernel (RBF) for classification. Overall, they achieved over 95\% accuracy for identifying drone sounds among other sounds in the test set.

The authors in \cite{jeon2017empirical} evaluated the effectiveness of several \ac{ML} models, such as Gaussian Mixture Model (GMM), Recurrent Neural Networks (RNN), and Convolutional Neural Networks (CNN), to detect and classify drone sounds using \ac{MFCC} features. They collected 64 seconds of raw drone sounds in a quiet outdoor place, using a single microphone located at distances of 30 meters, 70 meters, and 150 meters from the drone. They also collected 677 seconds of raw background sound from the ordinary real-life environment, with background noises such as airplanes, people chatting, and cars passing. In addition, they used publicly-available datasets mixed with the raw drone sounds to compensate for the lack of drone sounds in heterogeneous environments. The RNN model generated the best F-Score of 0.8009.

The authors in \cite{shi2018hidden} proposed a drones sound detection technique based on \ac{MFCC} features and using the hidden Markov model (HMM) tool for the classification. They tested their model on a database of several sounds, consisting of four drones, six cars, four planes, and four rain sounds. They organized the extracted features in five clusters, achieving an overall classification accuracy of up to 100\%, depending on the clusters selected for training and testing. They also evaluated the effect of the \ac{AWGN} on the extracted features, and they show that the classification performance is more than 80\% at \ac{SNR} of 5 dB, while it reaches the of 100\% at \ac{SNR} of 16 dB onwards.

The authors in \cite{kim2018neural} designed a real-time \ac{UAV} sound detection system based on artificial neural networks. The proposed system used the \emph{UrbanSound8K} dataset for non-drone existing training sounds, including general sounds such as air conditioner, car horn, children playing, dog bark, drilling, and motors idling, to name a few. The system applies the \ac{FFT} tool on the real-time sampled data and detects drones using Plotted Image Machine Learning (PIL) on the visualized \ac{FFT} graph, by comparing the average image similarity with a reference \ac{FFT} template associated with a target of interest. In addition, the system collects drones sound data from many reference points and merge them into a general comprehensive artificial neural network model, reporting the accuracy of 86.3\%. 

The authors in \cite{shi2018anti} designed ADS-ZJU, i.e., a hybrid anti-drone system that combines heterogeneous surveillance technologies, such as video, audio, and RF sensors. The acoustic signal features were extracted using \ac{STFT}, as well as reference features for video and RF signals. These values were used as the input of an \ac{SVM} algorithm for drone detection. The detection accuracy achieved using only acoustics signals ranges from 21.1\% at 100m to 87.4\% at 20m distance from the drone.

The sound used by a drone has been also used for tracking and localization purposes. In this context, the authors in \cite{chang2018surveillance} designed an acoustic system made up of two tetrahedron-shape acoustic arrays, aimed at collecting sounds periodically from the surrounding environment. The sounds recorded through the system are processed according to the \ac{TDoA} estimation algorithm and the Gauss a-priori probability density function (GPDF), to perform drones localization. The range of the system is around 100m.

The authors in \cite{benyamin2014acoustic} used a small tetrahedral microphone array to collect acoustic signals from class I \ac{UAS} (characterized by a Maximum Gross Takeoff Weight (MGTOW) of less than 9 kg). These sounds are used to detect the presence of a drone, leveraging an adaptive Kalman filter for drone tracking and a beamforming algorithm for drone detection. They achieved a detection accuracy of 99.5\% when using a pass-band filter of 800-1700 Hz for ranges below 600 meters, reporting a 3\% false alarm rate.

The authors in \cite{yue2018software} developed a drone detection and position estimation system based on a wireless acoustic sensor network, using the \ac{SVM} algorithm for classification, and commercial \ac{SDR} coupled with \ac{ML} scheme to decode the UAV’s telemetry protocols. They used an array of seven microphones to collect acoustic signals so that the data from any of the four orthogonal microphones can estimate the drone position, using the \ac{TDoA} algorithm. They achieved a drone detection accuracy of 95\% using the SVM algorithm, leveraging the features extracted using the \ac{STFT}, and using the Principal Component Analysis (PCA) technique for dimension reduction.

The authors in \cite{ramesh2019sounduav} designed SoundUAV, i.e., a system to fingerprint drones based on their acoustic noise. They evaluated 54 motors and 11 drones of the same make and model with 99.48\% classification accuracy.

The authors in \cite{hauzenberger2015drone} combined the Linear Predictive Coding (LPC) and the zero-crossing rate (ZCR) techniques to develop a drone sound detection system. They recorded the sound emitted from some drones using three microphones, in an anechoic chamber with 30\%, 40\%, 50\%, 60\%, 70\%, 80\% and 90\% of the drones’ throttle level, and they extracted reference features. Then, they chose 50\% throttle level recordings - which is approximately the level when the drone starts to hover - to create a reference database. 
They achieved the most successful test results using the WLtoys V262 Cyclone Quadcopter (close to 100\% accuracy, using sounds recorded at a distance less than 20m), while the worst performances were achieved using the small Hubsan X4 (around 80\% classification accuracy, using sounds recorded at a distance less than 10m).

In the field, there are also a few commercial solutions. It is worth mentioning the Discovair G2 \cite{G2} system, which is a counter-UAS solution that utilizes an array of 128 interconnected microphones to establish a real-time azimuth and elevation of the target, using advanced digital signal processing. Similarly, DroneShield \cite{DS} is another solution that incorporates acoustic sensors in a multi-sensor framework, to create an anti-drone platform. Another solution is Ctrl+Sky \cite{CS}, which is a multi-sensor counter-drone system that can detect, track, and neutralize drones.

Table~\ref{tab:relatedcomp} summarizes the discussion above, cross-comparing the discussed studies across several system features, such as the used classifier, the features, the eventual consideration of the noise, the specific hardware, and the distance from the target. We notice that, at the time of this writing, none of the analyzed contributions investigated the impact of the payload the drone is carrying on the profile of the emitted sound. In addition, none of them provided hints on how to identify the weight the drone is carrying. Finally, note also that most of the previous work did not release the data used for the related analysis.

\begin{table*}[htbp!]
  \centering
  \begin{tabular}{|P{1.1cm}||P{1.8cm}|P{2cm}|P{2.5cm}|P{2.1cm}|P{1.6cm}|P{2.05cm}|}
\hline
\textbf{Ref.} & \textbf{Classifier} & \textbf{Features} & \textbf{Noise} & \textbf{Hardware} & \textbf{Distance [m]} & \textbf{Weight Detection}\\ \hline
\cite{anwar2019machine} &  Cubic SVM & MFCC & \begin{tabular}[c]{@{}c@{}}real\\ environment\end{tabular} & microphones array & - & \xmark \\ \hline
\cite{shi2018hidden} & HMM & MFCC & AWGN & - & - & \xmark \\ \hline
\cite{bernardini2017drone} & RBF SVM & STE, ZCR, TC, SC, SRO, MFCC & \begin{tabular}[c]{@{}c@{}}real \\ environment\end{tabular} & Audio sensor & - & \xmark \\ \hline
\cite{jeon2017empirical} & GMM, RNN, CNN & MFCC & quiet outdoor  place + added noise & single microphone & 30, 70, 150 & \xmark \\ \hline
\cite{kim2018neural} & \begin{tabular}[c]{@{}c@{}}PIL, ANN\end{tabular} & FFT & \begin{tabular}[c]{@{}c@{}}real\\ environment\end{tabular} & Low cost microphones & - & \xmark \\ \hline
\cite{shi2018anti} & SVM & \begin{tabular}[c]{@{}c@{}}STFT\end{tabular} & \begin{tabular}[c]{@{}c@{}}real\\ environment\end{tabular} & 4-Mics array & 20-100 & \xmark \\ \hline
\cite{chang2018surveillance} & \ac{TDoA} & GPDF & \begin{tabular}[c]{@{}c@{}}real \\ environment\end{tabular} & Two tetrahedral arrays & 100 & \xmark \\ \hline
\cite{benyamin2014acoustic} & Threshold & beam forming algorithm & \begin{tabular}[c]{@{}c@{}}real \\ environment\end{tabular} & Tetrahedral array & 600 & \xmark \\ \hline
\cite{yue2018software} & SVM & \begin{tabular}[c]{@{}c@{}} STFT\end{tabular} & \begin{tabular}[c]{@{}c@{}}real\\ environment\end{tabular} & 7  mics array & - & \xmark \\ \hline
\cite{ramesh2019sounduav} & RBF SVM & Cepstral features & \begin{tabular}[c]{@{}c@{}}Lab\\ environment\end{tabular} & Blue Yeti Pro microphone & - & \xmark \\ \hline
\cite{hauzenberger2015drone} & Threshold & \begin{tabular}[c]{@{}c@{}}LPC, ZCR\end{tabular} & \begin{tabular}[c]{@{}c@{}}anechoic\\ chamber\end{tabular} & 3 microphones & 2-40 & \xmark \\ \hline
Our paper & Cubic SVM & MFCC & real environment and AWGN & Rode VideoMic Pro & 7 & 0-500g \\ \hline
\end{tabular}
  \caption{Overview and Comparison of related contributions on drones acoustic analysis.}
  \label{tab:relatedcomp}
\end{table*}

%
%
%

\section{Reference Scenarios}
\label{sec:scenario}

\begin{figure}[htbp]
    \centerline{\includegraphics[width=\columnwidth]{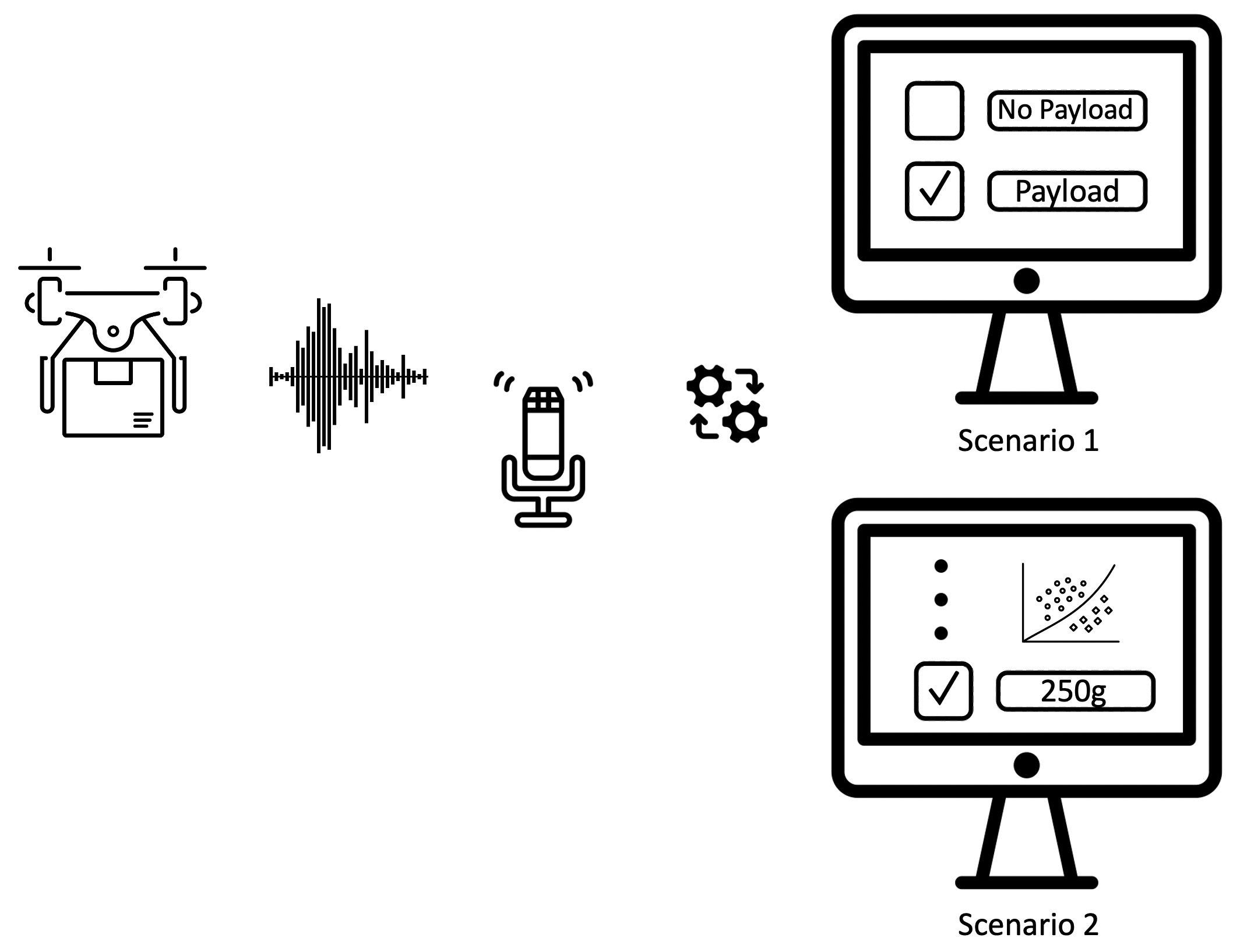}}
    \caption{Reference Scenarios considered in our work.}
    \label{fig:scenarios}
\end{figure}

In this paper, we considered two reference scenarios, illustrated in Figure~\ref{fig:scenarios} and discussed below.

{\bf Scenario 1.} In this scenario, we assume that a microphone is deployed in a remote location, e.g., a desert or an agriculture field. Due to the nature of the scenario, we assume that it is unfeasible to deploy a wireless camera to allow a remote administrator to immediately visualize the drone. The microphone is connected to a constrained device, such as an \ac{IoT} device, that has limited computational capabilities. Given the reduced capabilities of the computing device, we assign it a lightweight objective: to detect if the drone is carrying a payload, or not.
Further, we assume that other systems, already available in the literature, are deployed in conjunction with our system both to identify the type of the drone via its audio recordings (e.g., \cite{ramesh2019sounduav}) and to identify if the drone is hovering or moving (e.g., \cite{sciancalepore2019_wiseml}). 
As we will show later on, the IoT device can record the drone sound for a limited time, and then analyze the pitch of the recorded sound to identify if the drone is carrying a payload or not.


{\bf Scenario 2.} In this scenario, we consider a collaborative use-case, where a user would like to determine if a specific well-known drone is carrying a specific payload. Specifically, we assume that a user is receiving a specific package via a drone-based delivery system, such as the ones already in place by commercial entities, such as Amazon \cite{bamburry2015drones} \cite{d2014guest}. Despite a manual inspection of the package could be preferable to identify if the delivery is as expected, often opening a package could compromise the possibility for the user to ask for a refund or replacement of the item. Therefore, without touching the drone nor the package the drone is carrying, but only by analyzing the sound emitted by the drone, the user would like to verify that the weight of the payload the drone is carrying is compatible with the one he/she is expecting. Note that this is a more sophisticated scenario, where it is required to distinguish finer payload differences---we set the bar to 50 grams. Therefore, we assume that the drone hovers in front of the house of the customer for a limited time, where there is a microphone, directly connected to a standard laptop. On the laptop, the customer has installed a software tool that can analyze the sound emitted by the drone and identify if the payload the drone is carrying is different from the expected one. To achieve its objective, the software tool can leverage a pre-trained machine learning model, where the model of the drone is known in advance---this tool could be provided by the seller or by a trusted third party.

We highlight that the two scenarios described above are only a reference, and our research can be directly applicable also to other use-cases. For instance, surveillance towers around critical infrastructures can deploy the proposed system to get an estimation of the weight of the payload a drone is carrying. Indeed, having an estimation of an unauthorized drone weight and payload can be beneficial in determining the best countermeasure plan. Another use-case applies to military bases, where the drone has to hover in-place before entering the base, to examine its sound fingerprint and verify that the drone is not carrying an extra payload compared to the expected one.

Finally, we emphasize that, in all the use-cases discussed above, we assume that the drone is not moving, but hovering at a given distance from the microphone. The reason for this choice will be clearer later on and is mainly due to the hardly-predictable patterns that the sound produces when activating its motors for moving. 
On the one hand, we emphasize that considering a hovering drone is pretty reasonable for the real-life use-cases identified above, where this requirement does not represent a serious limitation. On the other hand, we highlight that, to the best of our knowledge, this study is the first one to analyze the relationship existing between the sound a drone emits and the payload it is carrying, and analyzing the modification of the sound of the drone when the drone is moving is left as future work, to further boost the research efforts in the UAV domain.

\section{Experimental Setup}
\label{sec:setup}

In this section, we illustrate the equipment used for the acquisition and processing of the sound emitted from an \ac{UAV}.

\begin{figure}[!htbp]
    \centerline{\includegraphics[width=\linewidth]{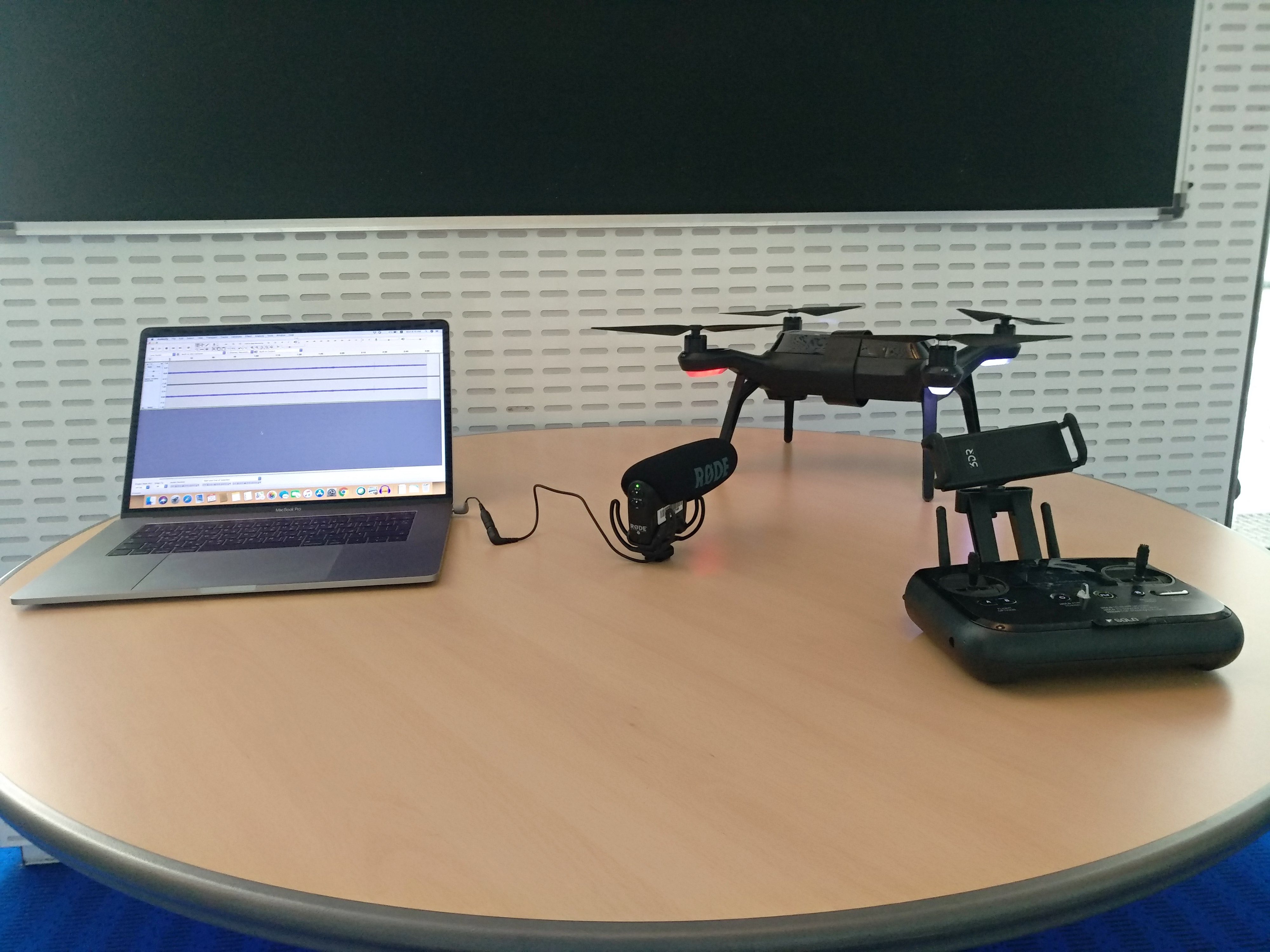}}
    \caption{Equipment setup for the live measurements of the sound emitted from the drone.}
    \label{fig:equipment}
\end{figure}

Figure~\ref{fig:equipment} shows the setup we used for our experiments.

The details of our equipment are reported in the following.

\textbf{Drone.} We selected the 3DR SOLO drone~\cite{3DRmanual}. It is a medium-sized commercial drone, weighing about 1.5 kg, including the 14.8 V battery (approx. 0.5 kg).
The drone can carry a maximum payload weight of 700g, including the 3DR Gimbal and a GoPro, weighting approximately 390g \cite{3drpayload}. In our experiments, we did not include these optional elements, and thus we tested payloads up to a weight of 500g. We also remark that we used the drone as it is out of the box, without any hardware or software modification. The 3DR Solo drone is characterized by an open-source architecture, featuring the Pixhawk 2.0 flight controller and the ArduCopter 3.3 firmware, currently adopted by more than 30 commercial drones~\cite{sciancalepore2019_wiseml}.

\textbf{Microphone.} To record the sound emitted from the drone, we used the  Rode VideoMic Pro directional microphone \cite{rode}. It is a battery-powered directional microphone, equipped with embedded filters and level controls. The native bandwidth of the microphone includes the range $\left[ 40Hz, 20kHz \right]$, and low-bandwidth noises such as air conditioners and traffic noises could be automatically rejected thanks to an optional high-pass filter at 80Hz. It also includes a $-10$~dB level attenuator, ideal for recording loud or very-close noise sources. The connection to the host is provided using a 3.5mm stereo mini-jack output (dual mono) \cite{rode}.

\textbf{Laptop.} For the analysis of the sounds recorded from the drone, we used a MacBook Pro laptop, equipped with a 2.9 GHz Intel Core i7 processor and 16 GB of RAM.

\textbf{Mission Planner software tool.} \emph{Mission Planner} is a software tool used to interface with the 3DR Solo drone, to get telemetry data, and to instruct the drone to accomplish a specific mission. In this paper, we use this tool to get information about the rotation speed of the motors, and a few other pieces of information, such as the ones related to the stability of the drone.

\textbf{Measurements Collection.} To collect the measurements, we used the Audacity audio software tool \cite{team2007audacity}. It is an easy-to-use, multi-track audio editor, that allows to easily select the parameters for the audio collection and the configuration of the microphone. Specifically, we notice that all the sounds used for this work were collected at the sampling rate of $44.1$ KHz, to maintain a good recording quality over the microphone's 20 kHz frequency bandwidth, in accordance to existing works \cite{shi2018hidden}, \cite{ramesh2019sounduav}. 

\noindent
The drone sound recordings were collected in an open, outdoor environment, placing the microphone at a distance of $7$ meters from the hovering drone. For each measurement, lasting $170$ seconds, the weight payload that the drone was carrying has been changed. We experimented with a total of $11$ different payload weights, ranging from $0$ g to $500$ g, with steps of $50$ g. We remark that each recording starts at the time the drone is hovering, and it ends before the drone lands, to capture only the sound of the drone hovering in place. The use-cases corresponding to this specific behavior of the drone are consistent with the scenarios described in Section~\ref{sec:scenario}. All the data gathered for this study have been publicly-released as open-source~\cite{data}.

\noindent
The output of audacity is a \emph{WAV} file, that is provided as input to the Matlab 2019b software tool for analysis.

\section{Sound Components Analysis}
\label{sec:sounds}

One of the most diffused metrics to analyze a generic sound is the \emph{pitch}. The \emph{pitch} of a sound is defined as the fundamental frequency of an audio signal~\cite{caprolu2020_comst}. For instance, in the context of speech analysis, the pitch is often used to discriminate male from female voices, or to identify the specific speaker. Overall, identifying the pitch of a signal is not straightforward, as natural sounds arise as the combination of different frequency components, alternating in period and/or amplitude~\cite{sun2002pitch}. 

Considering the sound emitted by a generic drone, several components contribute to defining the pitch of the emitted sound. In line with related works on sound analysis, we found that two components are mainly contributing to the sound: motors and blades \cite{djurek2020analysis}.

First, the rotation speed of the motors, measured in \ac{RPM}, contributes significantly to the sound emitted by the drone. Specifically, a drone continuously monitors and adjusts its motors rotation speed, to keep stability and compensate external factors. We highlight that the rotation speed of the motor is not fixed, but oscillates in a given interval. When the drone is hovering, the Electronic Speed Controller (ESC) component functions to maintain a steady rotation of the motors, to preserve the drone stability and adjust its position. If the drone is moving, the oscillation of the motors rounds depends on its speed, wind conditions, and other factors related to its stability, such as the compensation of a particular movement.
In addition, the blades contribute to the sound emitted by the drone. The blades determine the efficiency and smoothness of the flight, and their presence is key to allow the drone to fly.

\begin{figure}[!htb]
\centerline{\includegraphics[width=\linewidth]{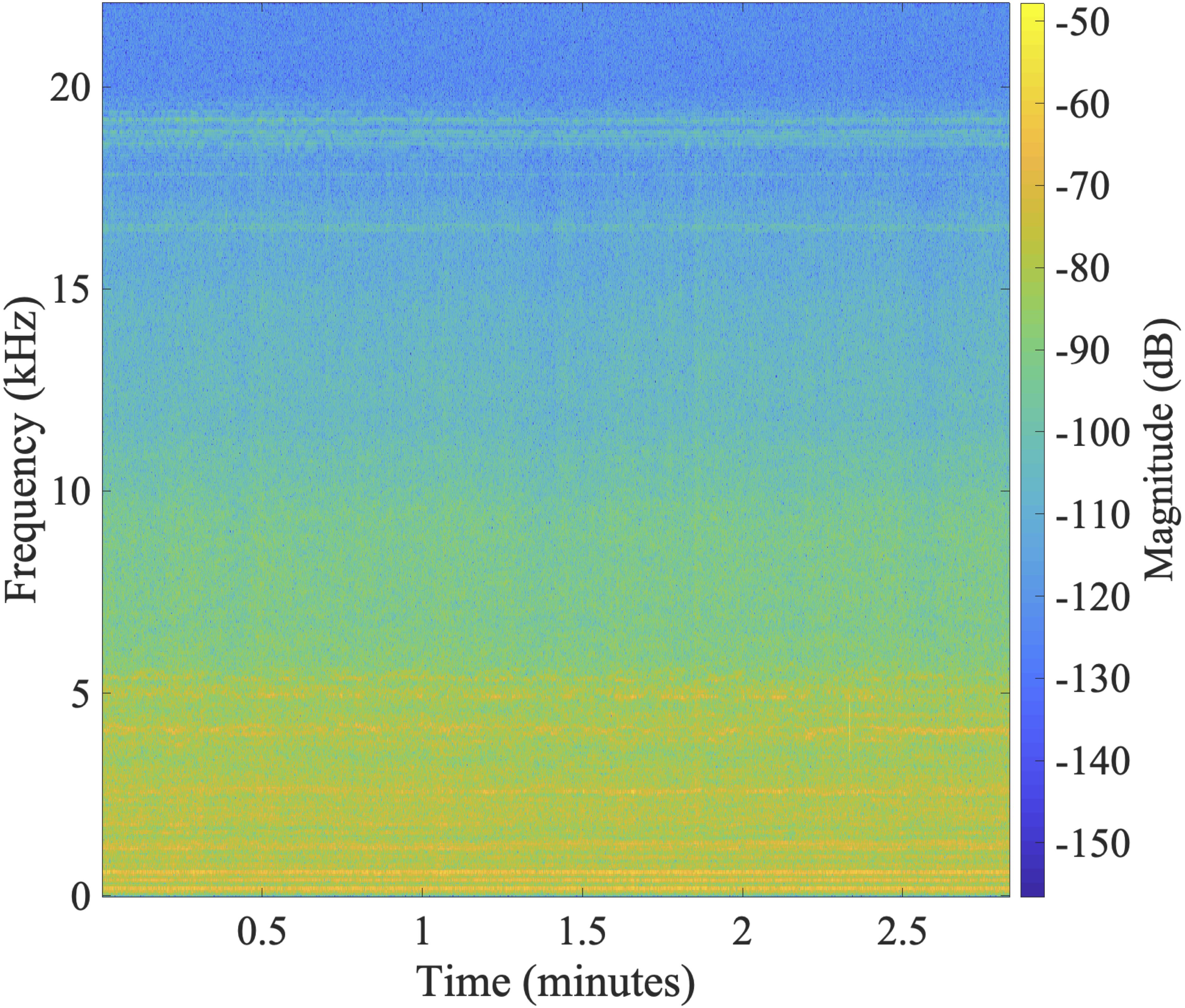}}
\caption{Full Spectrogram of drone sound, when no payload is carried, over 170 seconds. Lighter colors correspond to stronger frequency components.}
\label{fig:specFull0gPitch}
\end{figure}

To show the variation of the pitch during a specific interval, in Figure~\ref{fig:specFull0gPitch}, we report the full spectrogram of the sound recorded from the 3DR solo drone at the reference distance of 7 meters from the microphone, when no payload is carried by the drone.

We can see that the strongest components, i.e., the frequency components characterized by higher sound emissions, are located in the low-frequency range, below $7$~kHz. In line with the findings of the authors in~\cite{djurek2020analysis}, we identify these frequency components to be associated with the blades of the drone.

To get more insights about the average speed of the motors when the drone carries different payload weights, we leveraged the telemetry data provided by the Mission Planner tool.

\begin{figure}[htbp]
    \centerline{\includegraphics[width=\linewidth]{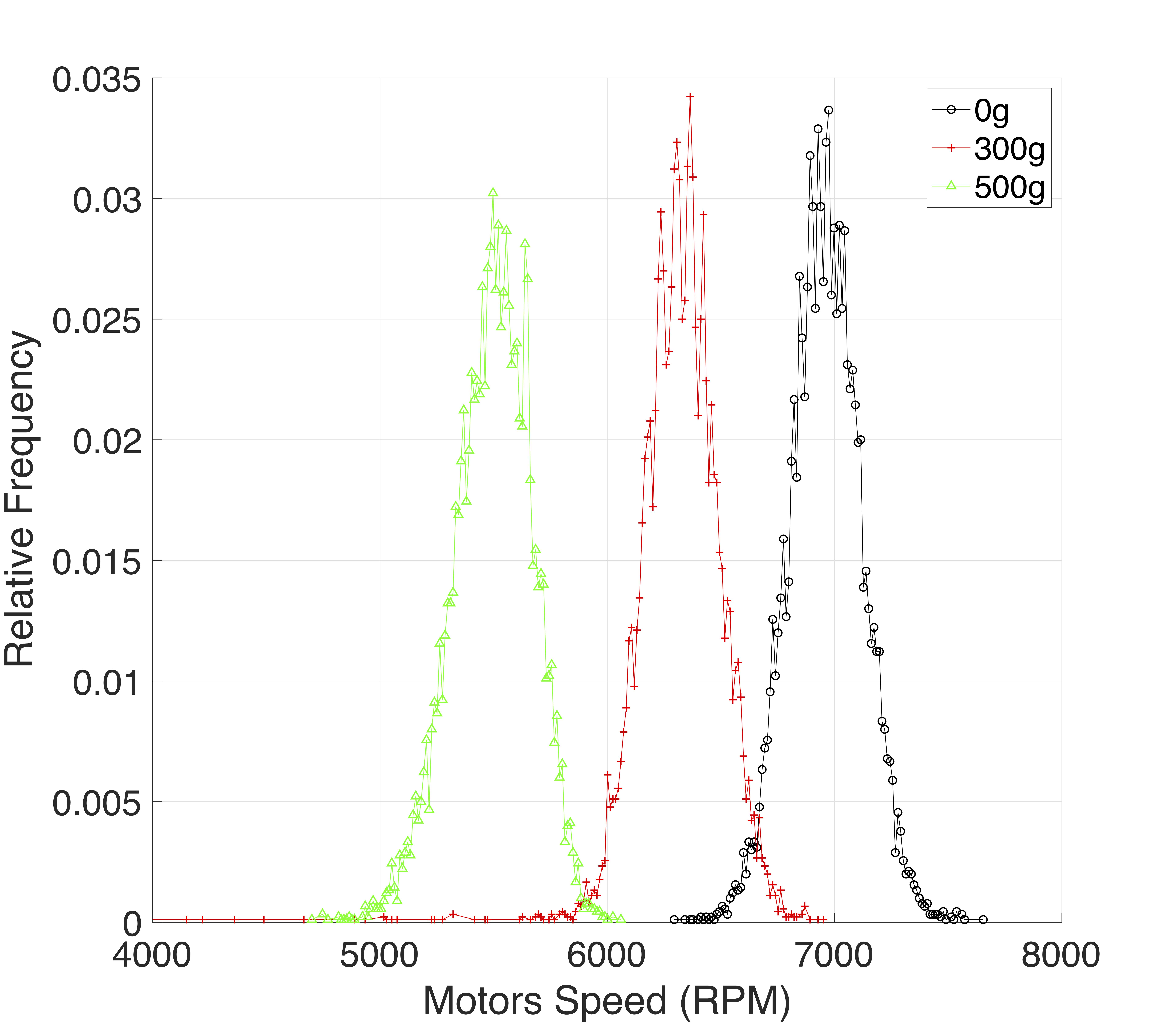}}
    \caption{Motors average speed distribution at different payload weights}
    \label{fig:RPMhist}
\end{figure}

Figure~\ref{fig:RPMhist} shows the histogram of the average drone motors RPM, as extracted from Mission Planner software tool, and their relationship to three different reference payload weights carried by the drone, i.e., $0$~g, $300$~g, and $500$~g, respectively.

First, we notice that for a particular payload weight carried by the drone the motors average speed is not constant. As previously highlighted, this is due to the continuous adjustments that the drone performs on the speed of the motors, to compensate instantaneous drops or movements, due to the pressure, inertia, and other environmental conditions (e.g., wind).

We also notice a clear difference between the recordings for different weights. Indeed, the motors of the drone have to rotate faster when the weight of the payload is higher, for the drone to hover and stabilize its position. 

In the \emph{Scenario 1} considered in this paper, we will focus on the pitch of the signal, and we will consider both the blades sound and the motors sound, even if the blades sound is the most prominent component. 
More details will be provided in Section~\ref{sec:scenario1}.

At the same time, from Figure~\ref{fig:specFull0gPitch}, we can see that there are several harmonics for the pitch, mainly located below the frequency of $7$~kHz. To take into account these components, we can use the \acl{MFCC} (MFCC) tool. The \ac{MFCC} is a technique to describe the different frequency components of the signal, and it is often used to best describe how the human phonetic system perceives the audio signals. 

\begin{figure}[htbp]
\centerline{\includegraphics[width=\linewidth]{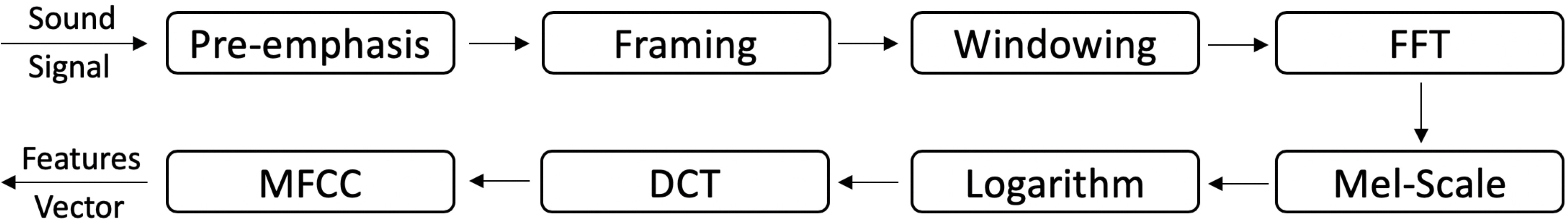}}
\caption{Extraction of the MFCC components of an audio signal.}
\label{fig:mfcc}
\end{figure}

Figure~\ref{fig:mfcc} depicts the features extraction steps using the MFCC technique \cite{rabiner2011theory}.

The input to the process is the digital representation of the audio signal, as it is provided from the microphone. A preliminary pre-emphasis stage is applied to enhance the energy of the higher frequencies of the sound and to improve the related \acl{SNR}. Then, the MFCC divides the recorded window in very short time windows, where the features of the input can be assumed as stationary. Later, the technique computes the \ac{FFT} of each window, and shifts the components to an alternative frequency axis, referred to as the non-linear Mel scale, further divided into a set of bands, known as the Mel bands. 
Specifically, the content of each Mel band is computed according to Eq.~\ref{eq:mel}.
\begin{equation}
\label{eq:mel}
    Mel(f) = 2595 \cdot \log_{10}\left(1 + \frac{f}{700}\right).
\end{equation}

Then, the technique envisions the computation of the logarithm of the Mel spectrum (referred to as $D(m)$), and finally, the computation of the cepstral coefficients $c(n)$ using the discrete cosine transform (DCT), according to the following Eq.~\ref{eq:cepstral}. 

\begin{equation}
\label{eq:cepstral}
    c_n =
     \sum_{m=0}^{M-1} D(m) \cos \left[\frac{\pi n (m-0.5)}{M} \right], \left (n = 0, \dots, R \right),
\end{equation}

where R is the total number of MFCC coefficients and the index $m$ identifies the specific Mel-band. Note that the MFCC components consider the input frequency range $\left[633, 6,854\right]$~Hz, thus being perfectly applicable to our problem.

\begin{figure}[htbp]
     \centering
     \includegraphics[width=\columnwidth]{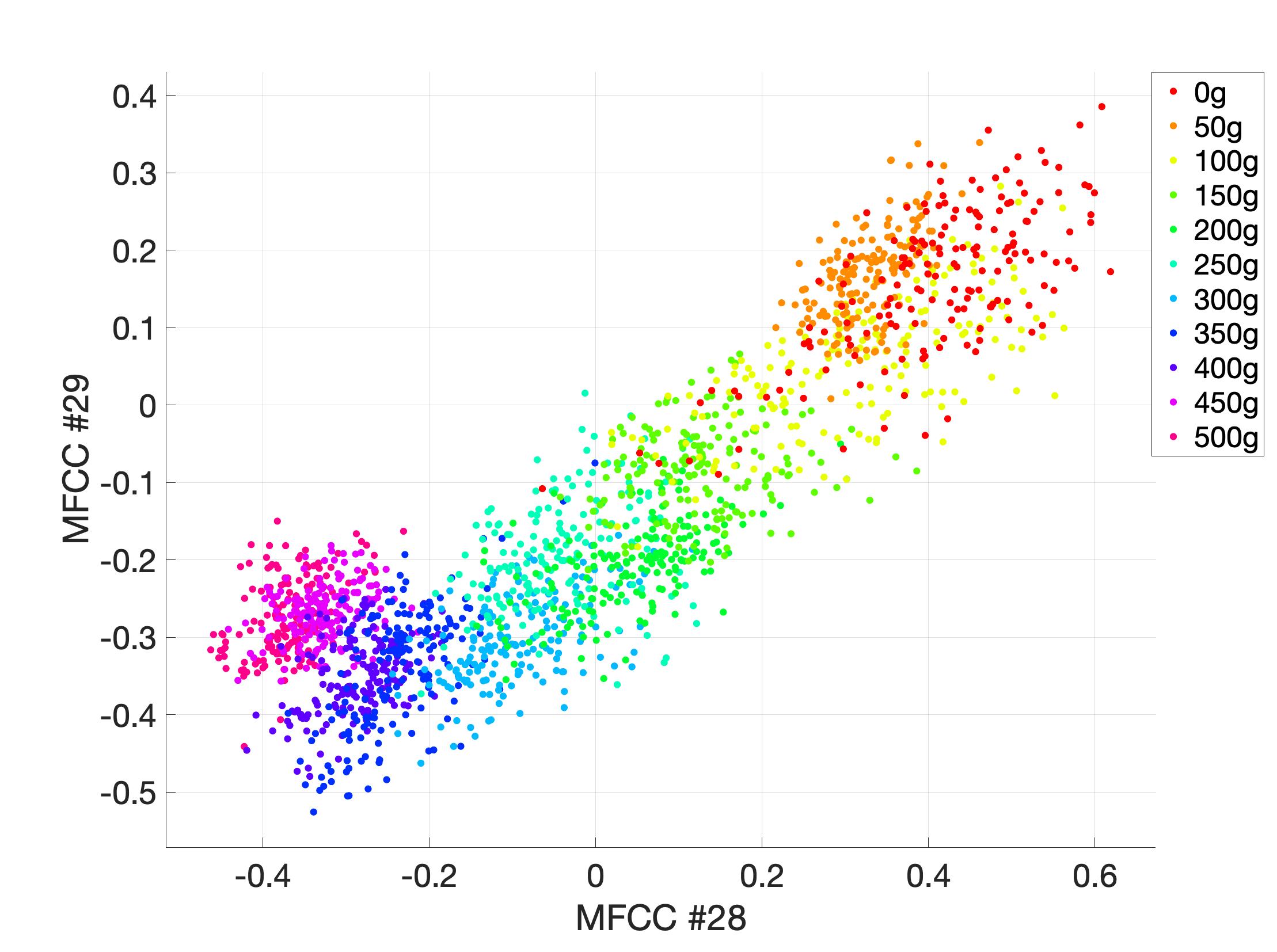}
     \caption{Relationship between the MFCC components $28$ and $29$, extracted from the recordings of the sound of a drone carrying different weights, from $0$~g to $500$~g, considering a time window of $1$~s.}
     \label{fig:ref_MFCC28vs291s}
\end{figure}

As a reference example for the value of the MFCC features in the analysis of the weight carried by the drone, in Figure~\ref{fig:ref_MFCC28vs291s}, we show the relationship existing between two reference MFCC components of the signals recorded when the reference drone carries payloads with different weights, i.e., the 28th (bandwidth $\left[ 2,618, 3,004 \right]$~Hz) and the 29th (bandwidth $\left[ 2,804, 3,217 \right]$~Hz). Note that the MFCC components showed in the figure have been computed considering the reference time windows of $1$~s.

We can notice that there is a clear trend in the relationship between the showed MFCC components. First, we notice that similar weights produce points that are very close to each other, creating well-defined clusters. 
At the same time, we can notice that, with the increase in the weight carried by the drone, the cluster moves down across a line, parallel to the reference line $y=x$, indicating a clear correlation between the weight carried by the drone and the produced points. 

We remark that Figure~\ref{fig:ref_MFCC28vs291s} shows the relationship only between two of the several available MFCC components. 
In Section~\ref{sec:results} we will show how the MFCC of the signals can be used to discriminate the specific weight of the payload a drone is carrying.

\section{Experimental Analysis and Results}
\label{sec:results}

In this section, we report the results of our investigation about the capability to detect the weight of the payload a drone is carrying, based on the analysis of the sound emitted by the drone. Specifically, Section~\ref{sec:scenario1} illustrates the results for the \emph{Scenario 1} described before, while the results for the \emph{Scenario 2} are reported in Section~\ref{sec:scenario2}.

\subsection{Scenario 1}
\label{sec:scenario1}

We recall that the Scenario 1 introduced in Section~\ref{sec:scenario} involves devices that do not have extensive computational capabilities, and therefore they can rely on lightweight operations only. The objective, in this case, is to ascertain whether the drone is carrying a payload. 

Based on the analysis carried out in Section~\ref{sec:sounds}, we identify the \emph{pitch} of the sound emitted by the drone as a feasible tool to accomplish such an objective.

Specifically, based on our observations and the values reported by the Mission Planner software tool, we identified that the pitch of the sound recorded by the drone is always in the range $\left[180-235 \right]$~Hz, despite instantaneous noise components can distort its smooth identification on the receiving microphone. Therefore, on the sound recorded by the microphone, we apply a band-pass filter in the frequency range $\left[180-235 \right]$~Hz, to filter out noise components.

\begin{figure}[htbp!]
\centerline{\includegraphics[width=\linewidth]{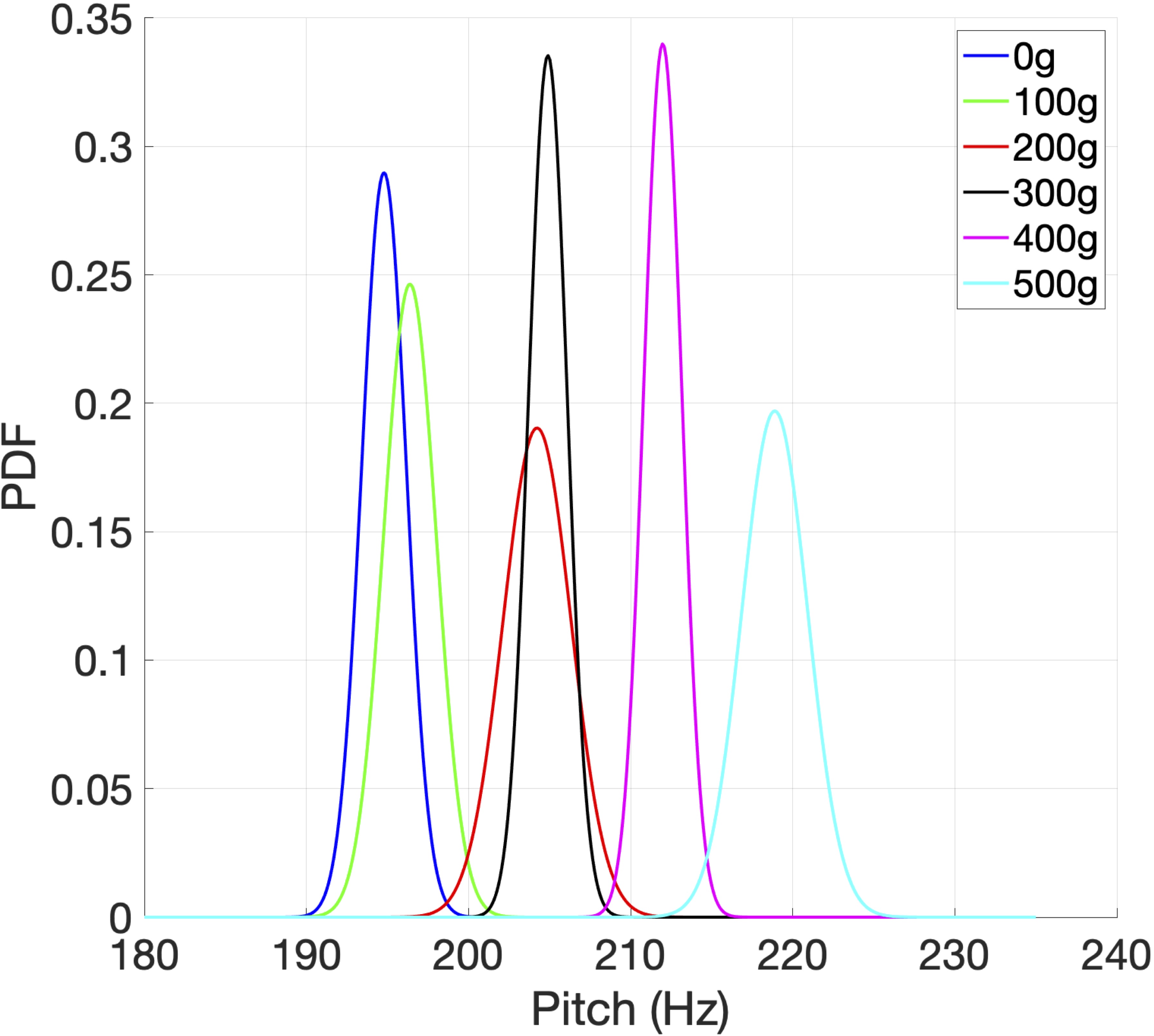}}
\caption{Probability distribution function associated to the values of the pitch, recorded for different payload weights and computed over a reference time windows of 2.5 seconds.}
\label{fig:pitchPDF}
\end{figure}

We recorded the drone sound for $11$ different weights, ranging from $0$~g to $500$~g, with $50$~g step, for a duration of $170$~s each. Then, we divided each trace into shorter time-slots, each lasting $2.5$~s, obtaining $68$ reference time blocks for each recording. 
Figure~\ref{fig:pitchPDF} reports the probability distribution function of the pitch for the sound generated by the drone when the drone is hovering while carrying different payload weights. 
For the sake of clarity, we show only $6$ reference distributions.

In line with the content of the previous Figure~\ref{fig:RPMhist}, we notice that increasing the payload weight leads to an increase in the mean value of the associated pitch frequency. 
This is due to the frequent oscillations and adjustments performed by the motors and the blades of the drone, leading to a dispersed profile around the mean pitch value. We also note that the variance of the extracted probability density functions is almost the same for all the recordings, but the one for the higher payload weight of $500$~g. Our intuition is that the drone oscillates more with payload weights close to the maximum, and thus it requires more thrust to compensate for the turbulence experienced while hovering.

We also notice that similar payload weights are characterized by partially overlapping probability distribution functions, making it hard to detect the specific weight carried by the drone by looking only at the pitch of the sound. On the one hand, we highlight that the recordings have been obtained using a low-end directional microphone, that has not been designed specifically for highly-precision tasks. Thus, we expect that more performing microphones should be able to remove noise components, further enhancing the identification of the pitch. On the other hand, we notice that the higher the payload weight carried by the drone, the lower the probability that the profile of the distribution function of this high payload weight overlaps with the one corresponding to $0$~g, i.e., the sound of the drone when no payload is carried.

\begin{figure}[htbp]
\centerline{\includegraphics[width=\linewidth]{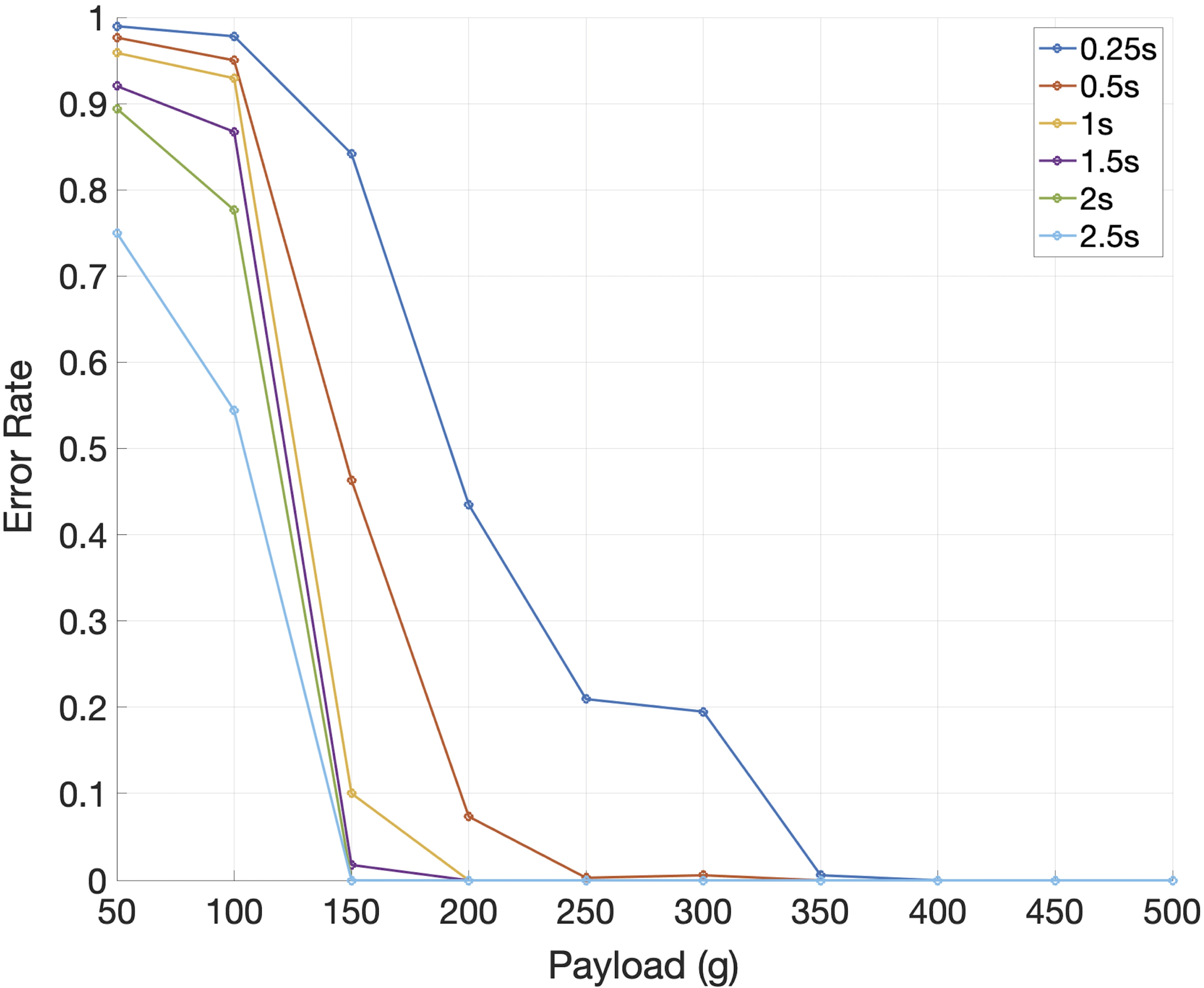}}
\caption{Error Rate of discriminating if the drone is carrying a payload or not, by considering different weights and increasing time windows.}
\label{fig:errorVswindow}
\end{figure}

To show the capability to discern whether a drone is carrying a payload via the pitch of the sound only, we considered each of the tested weights, and we computed the pitch over an increasing time window, from $0.25$~s to $2.5$~s. Then, we evaluated how many samples from each distribution fall within the distribution corresponding to the $0$~g payload weight (i.e., the error rate). The results are reported in Figure~\ref{fig:errorVswindow}.

We notice the error rate in the identification of a payload carried by the drone depends both on the specific weight carried by the drone and by the size of the time window, i.e., the acquisition time of the recording. Considering a specific time window, we notice that the error rate decreases with the increase of the weight carried by the drone. Indeed, recalling Figure~\ref{fig:pitchPDF}, the probability that the distribution corresponding to a specific weight carried by the drone overlaps with the distribution corresponding to $0$~g decreases with the increase of the weight carried by the drone. At the same time, computing the pitch over a larger time frame helps to reduce spurious oscillations, due to the adjustments that the motors of the drone perform to keep stability and compensate external effects. For instance, considering the weight of $200$~g payload, while computing the pitch over a time-frame of $0.25$~s leads to an error rate of the $42$\%, computing the pitch over a time-frame of $2.5$~s leads to a negligible error rate, very close to $0$. We also notice that lighter weights, such as 50 or 100 g, are characterized by a high error rate, and the only way to mitigate this effect is to increase the time-frame of the recording. Depending on the particular application scenario, this could be challenging (e.g., adversarial scenarios, where the system does not have control of the drone).

\subsection{Scenario 2}
\label{sec:scenario2}

We recall that the Scenario 2 introduced in Section~\ref{sec:scenario} involves devices that do not have strict constraints in their computational capabilities, and therefore, they can rely on more processing-intensive operations. The objective, in this case, is to identify the specific weight of the payload that the drone is carrying.

Based on the analysis carried out in Section~\ref{sec:sounds}, we identify the \emph{MFCC} components of the sound emitted by the drone as a feasible tool to accomplish such an objective.

\begin{figure}[htbp]
    \centerline{\includegraphics[width=\linewidth]{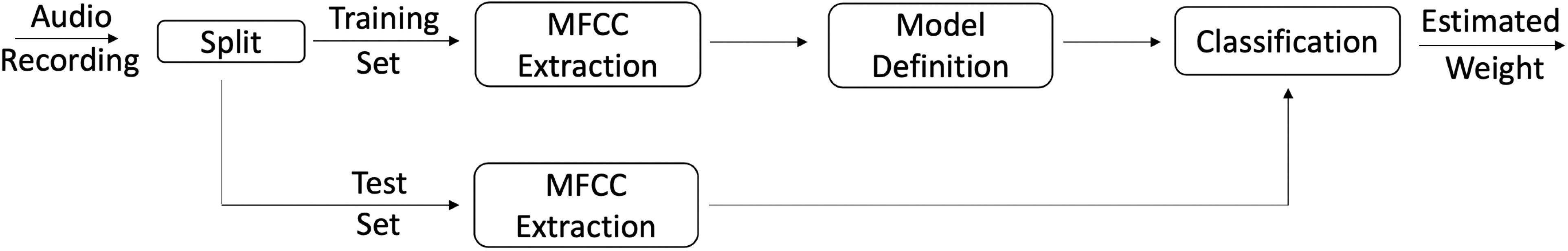}}
    \caption{Overview of the system set up to identify the weight carried by the drone via its acoustic emissions.}
    \label{fig:ml_system}
\end{figure}

\textbf{System Setup.} We set up a system composed of different stages, as shown in Figure~\ref{fig:ml_system}.

More in detail, for each of the collected recordings, we divided the trace into two parts. We adopt the first one as the \emph{Training Set}, consisting of the $70$\% of the size of the recording (approximately $120$~s), while the second part (\emph{Testing Set}), consists of the remaining $30$\% (approximately $50$~s) of the recording. For both sets, we obtained a spectral representation based on $40$ MFCC components, characterizing the content and the distribution of the sound within each bandwidth in the range of the microphone. 
Table~\ref{tab:mfccpara} summarizes the parameters used to generate the MFCC components.
\begin{table}[htbp]
    \centering
    \caption{Parameters used for the extraction of the MFCC components.}
    \label{tab:mfccpara}
    \begin{tabular}{cc}
    \hline
    \textbf{Parameter} & \textbf{Value} \\ \hline
    Sound sample length per instance & 0.25~s, 1~s  \\ \hline
    Sampling frequency, $f_{s}$ & 44.1 kHz  \\ \hline
    Window Size & $0.03f_{s} [s]$  \\ \hline
    Overlap Length & $0.02f_{s} [s]$  \\ \hline
    FFT Length & $0.03f_{s}$ [s] \\ \hline
    Number of features & 40  \\ \hline
    \end{tabular}
\end{table}

To show the effect of an increasing time window, we consider two different time windows for the MFCC features generation: $t_1 = 0.25$~s and $t_2 = 1$~s.

\begin{figure}[htbp]
    \centerline{\includegraphics[width=\linewidth]{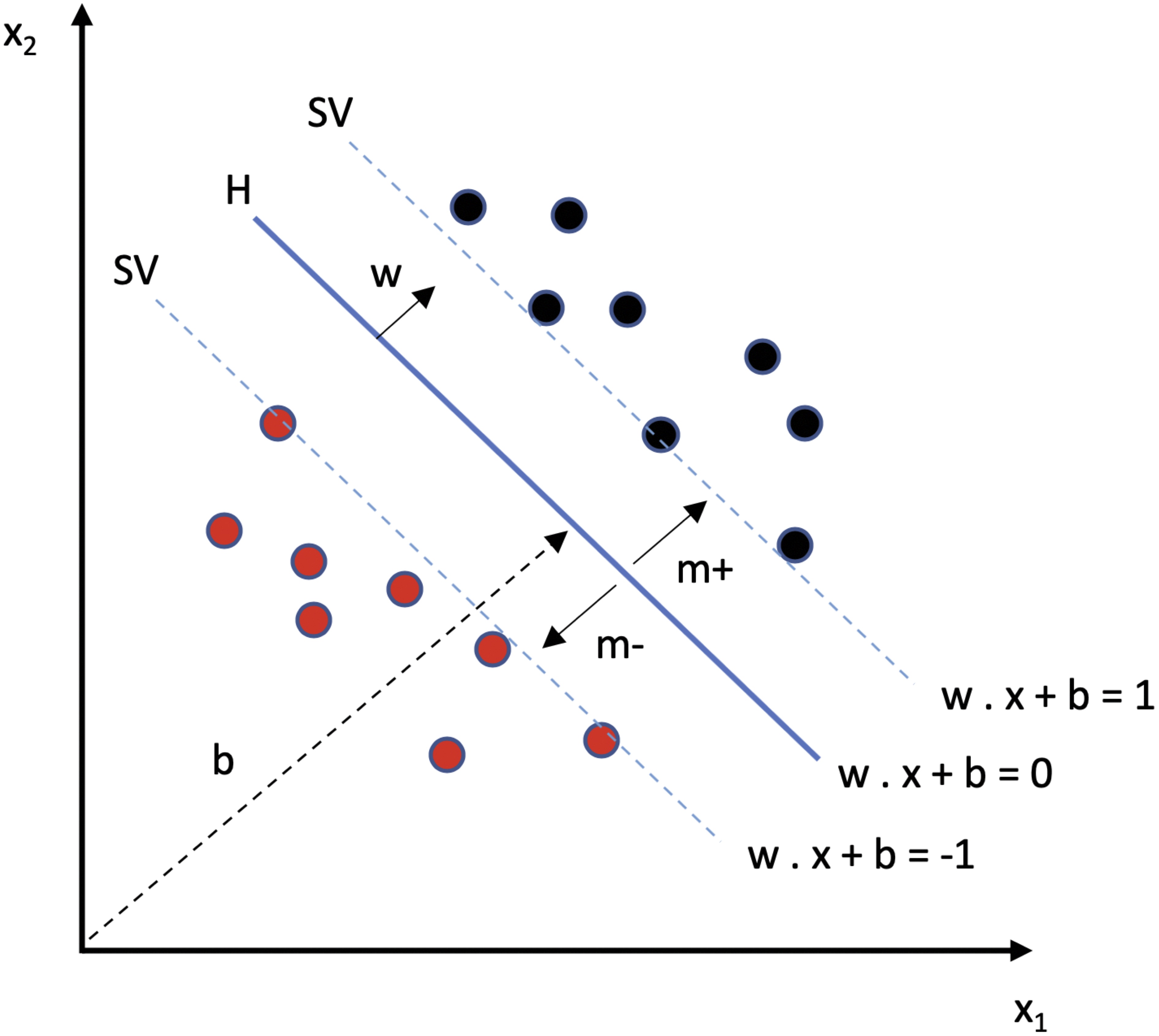}}
    \caption{Logic of the SVM classifier}
    \label{fig:svm}
\end{figure}

Then, for the classification task, we used the \acl{SVM} algorithm. 
To introduce the logic of the SVM, let us consider the toy example in Figure~\ref{fig:svm}. 

We recall that the \ac{SVM} technique is based on the creation of an optimal hyperplane ($H$), that acts as a decision boundary maximizing the margin of separation ($m$ = ($m+$) + ($m-$)) between the considered classes~\cite{boser1992training},~\cite{cortes1995support}. The training samples that are used to support the creation and the optimal location of the hyperplane are referred to as Support Vectors (SV), and they can be represented as per equation Eq.~\ref{eq:sv}.

\begin{equation}
\label{eq:sv}
    w^T\cdot x + b = \pm 1,
\end{equation}

where $w$ is the normal weight vector orthogonal to the hyperplane $H$, $x$ is the input vector, and $b$ is the bias. We highlight that, despite the SVM was designed as a binary classification tool, it can be generalized to perform multi-class classification, using techniques such as the \emph{One-vs-One} (adopted in this paper). 
Overall, assuming $N$ classes, the \emph{One-vs-One} technique trains a separate classifier for each different pair of classes in a dataset, generating 
$ {N \choose 2} = \frac{N(N-1)}{2}$ classifiers~\cite{bredensteiner1999multicategory}. A specific sample is assigned to the most feasible one of two classes, and the process is repeated considering that specific class and all the other classifiers. In the end, the class with the maximum number of \emph{votes} among all generated classifiers is considered as the reference one for the sample classification.

The capability of the decision boundary to separate different classes is improved when the input data are moved to a higher-dimensional space, through a process known as the \emph{kernel trick} \cite{kernel}. This technique enhances the capabilities of the SVM, allowing to classify non-linearly separable data. More in detail, the SVM polynomial kernel is defined as in the following Eq.~\ref{eq:kernel}.

\begin{equation}
\label{eq:kernel}
    K(x_1, x_2) = (x_1^T\cdot x_2 + 1)^p,
\end{equation}

where $p$ is the order of the polynomial. When $p=1$, the classification tool is referred to as \emph{linear SVM}, when $p=2$ it is a \emph{quadratic SVM}, when $p=3$ it is a \emph{cubic SVM}, and so on. The specific degree of the polynomial in Eq.~\ref{eq:kernel} should be chosen smartly, in a way to avoid over-fitting and adapt to the possible slight variations in the testing set. 

\begin{figure}
     \centering
     \includegraphics[width=\columnwidth]{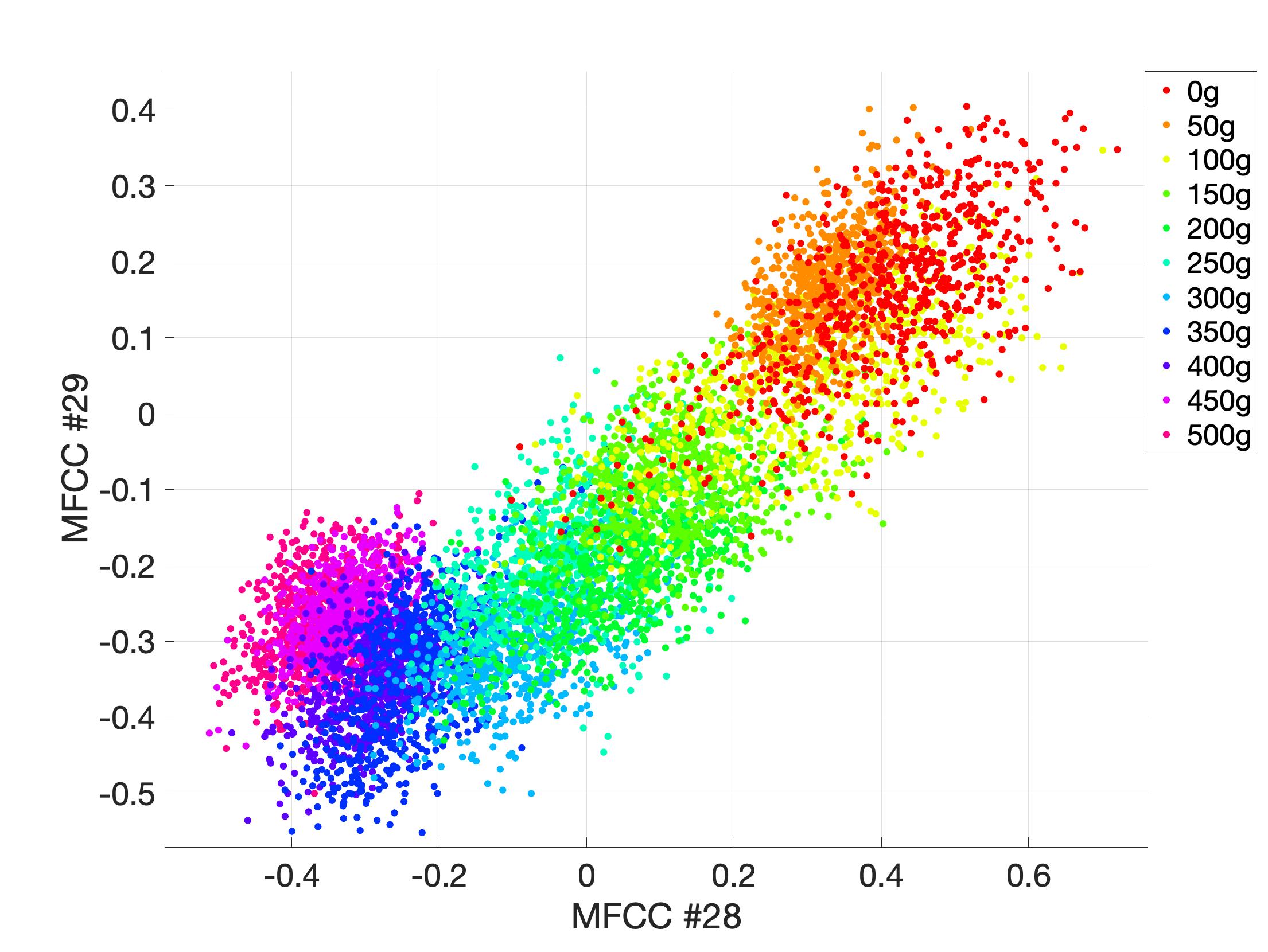}
     \caption{Relationship between the MFCC components $28$ and $29$, extracted from the recordings of the sound of a drone carrying different payload weights, from $0$~g to $500$~g, considering a time window of $0.25$~s.} 
     \label{fig:MFCCcomp}
\end{figure}

\textbf{Classification Results.} We implemented the system previously described using Matlab R2019a. In line with the analysis carried out in the previous subsection regarding the pitch, the size of the time window considered for the extraction of the MFCC components has an impact on the accuracy of the classification. To render this phenomenon, in Figure~\ref{fig:MFCCcomp} we report the relationship between two reference MFCC components, the 28th (bandwidth $\left[ 2,618, 3,004 \right]$~Hz) and the 29th (bandwidth $\left[ 2,804, 3,217 \right]$~Hz), when considering a time-window of $0.25$~s. We recall that the relationship between the same MFCC components for a time-window of $1$~s has been already reported in Figure~\ref{fig:ref_MFCC28vs291s}.

As previously highlighted, we notice the clear correlation existing between the points corresponding to the same weights. Indeed, they are located very close to each other, forming well-defined clusters. At the same time, note that increasing the weight by $50$~g leads to a shift of the cluster along a line that is mostly parallel to the bisect, further showing the clear relationship existing between the recorded sound and the specific payload weight carried by the drone. This behavior could also be beneficial to predict the sound expected for weights whose corresponding sound has not been recorded. An exception to this general behavior can be found for the samples of payload weights higher than $400$~g, and could be explained with the \emph{abnormal} activity performed by the motors and the blades to keep stability---needed when carrying a payload weight that is close to the maximum allowed one.

Furthermore, comparing Figure~\ref{fig:MFCCcomp} with  Figure~\ref{fig:ref_MFCC28vs291s} previously introduced, we notice that considering a shorter time window for the extraction of the MFCC components leads to an increasing overlap between neighboring payload weights. The reason is similar to the one generating the overlapping between the pitch of neighboring weights, and it is rooted in the fact that considering larger time windows mitigates transient noise effects, due to the continuous adjustments and compensations performed by the drone to keep a stable position against external environmental factors. At the same time, increasing the time window, i.e., increasing the recording time, increases also the probability to obtain different values of the MFCC components even for close payload weights, hence improving the
classification accuracy.

We remark that we showed the MFCC components $28$ and $29$ as they are among the most representative ones for the classification of the payload weight carried by the drone (two of the top four). However, other MFCC components are also relevant, and they further help in correctly classifying points that are located at the boundaries of the clusters showed in Figure~\ref{fig:MFCCcomp}. 
However, given the large number of available combinations for the whole set of $40$ MFCC components, we do not show them explicitly.

To justify the use of the SVM classification technique and to evaluate its feasibility for the payload weight acoustic recognition task, we applied several widely-used supervised ML classification algorithms to the MFCC components of the sound recorded when the drone carries different payload weights. They are: 1) Linear Discriminant (LD), 2) Quadratic Discriminant (QD), 3) Linear SVM (LSVM), 4) Quadratic SVM (QSVM), 5) Cubic SVM (CSVM), 6) Fine \ac{KNN} (FKNN), 7) Subspace KNN (SKNN), 8) Bagged Trees (BT), 9) Gaussian Naive Bayes (GNB), and, finally, 10) Kernel Naive Bayes (KNB). To cross-compare the performance of the different classifiers, we use the standard metric of the \emph{Classification Accuracy}, defined as in Eq.~\ref{eq:cl_acc}.
\begin{equation}
\label{eq:cl_acc}
    {Classification~Accuracy = \frac{ TP + TN }{TP + FP + TN + FN}} 
\end{equation}

where $TP$ refers to the True Positives, $TN$ to the True Negatives, $FP$ to the False Positives, and, finally, $FN$ to the False Negatives.

\begin{figure*}
     \centering
     \begin{subfigure}[b]{0.8\textwidth}
         \centering
         \includegraphics[width=\textwidth]{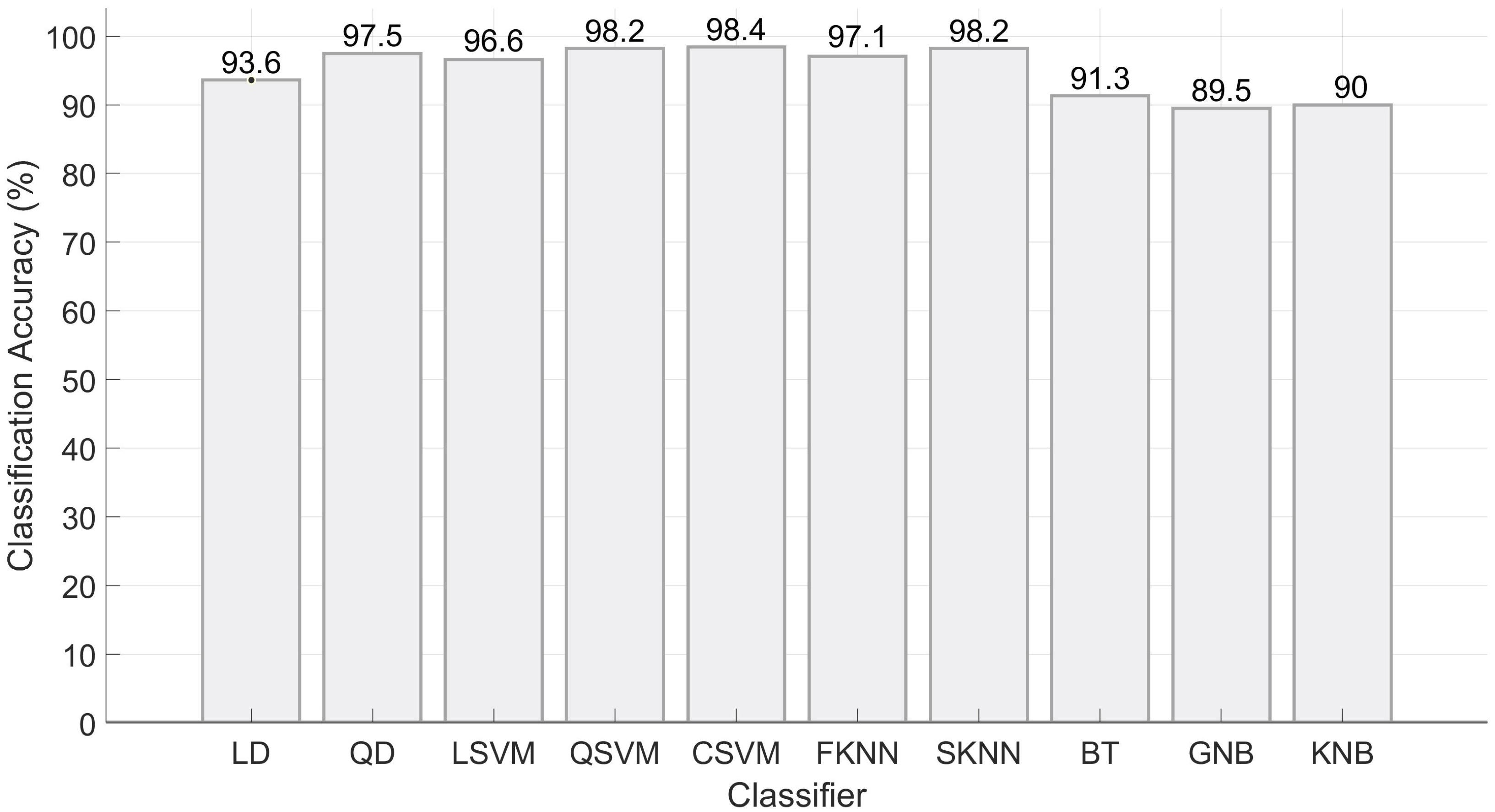}
         \caption{Performance using a time window $t=0.25$~s for the computation of the MFCC components.}
         \label{fig:MLdot25s}
     \end{subfigure}
     \hfill
     \begin{subfigure}[b]{0.9\textwidth}
         \centering
         \includegraphics[width=\textwidth]{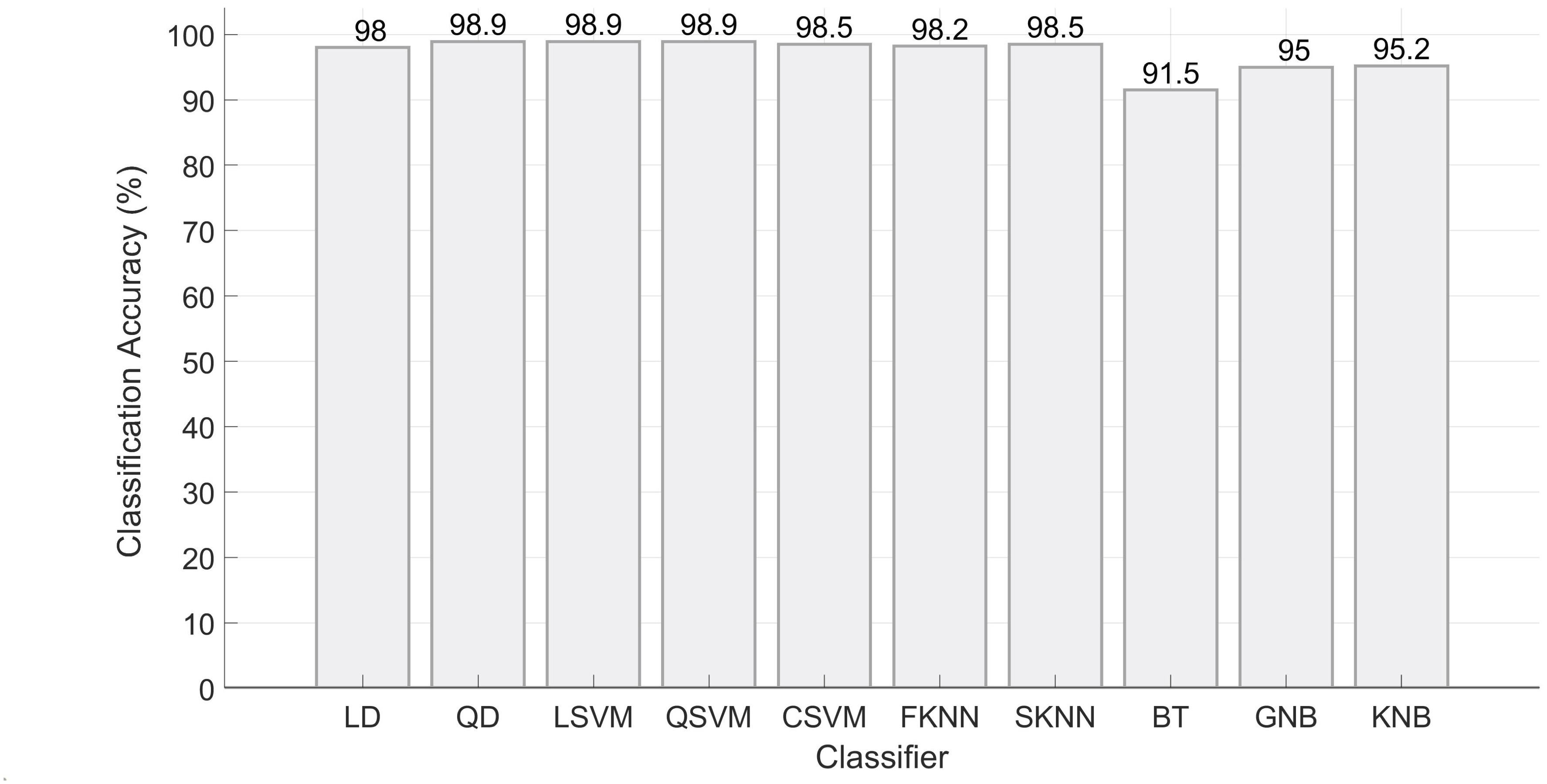}
         \caption{Performance using a time window $t=1$~s for the computation of the MFCC components.}
         \label{fig:ML1s}
     \end{subfigure}
     \caption{Classification Accuracy reported by several different supervised ML classifiers in the identification of the weight carried by the drone using its acoustic noise.}
     \label{fig:MLclass}
\end{figure*}

In Figure~\ref{fig:MLclass} we report the classification accuracy achieved by each of the above-introduced classifiers, considering a time-window of a) 0.25~s and b) 1~s for the computation of the MFCC components. We remark that the results have been obtained using the $70$\% of the samples for the \emph{training}, and the remaining $30$\% were used as the \emph{test set}. Similar results have been also achieved using the well-known cross-validation technique.

Figure~\ref{fig:MLclass} conveys a few interesting considerations. First of all, we notice that all the tested classifiers report classification accuracies over $89\%$ (the minimum classification accuracy is reported by the GNB classifier in Fig.~\ref{fig:MLdot25s}, and it is $89.5$\%). 
This finding confirms that, independently from the chosen algorithm to accomplish sound classification, there is a clear relationship between the sound emitted by the drone over the whole \emph{acoustic} frequency range and the payload weight carried by the drone. We remark that, at the time of this writing, to the best of the authors' knowledge, none of the contributions present in the literature highlighted this correlation. 

In addition, we notice that increasing the acquisition time from $0.25$~s to $1$~s improves the classification accuracy for all the tested classifiers. This finding further confirms the beneficial effect of increasing the observation time, to get rid of the noise components and the adjustments carried out by the drone to stabilize.

We also notice that several classifiers are achieving very high accuracy. For instance, looking at Figure~\ref{fig:ML1s}, the QD, LSVM, and QSVM all report a classification accuracy of $98.9$\%. However, comparing all the classifiers across the different recording times, we notice that the cubic SVM classifier is the one characterized by the least performance shift, i.e., $0.1$\%.

Without loss of generality, recalling our discussion about the classification logic of the SVM, we find that, independently from the degree of the kernel polynomial, the SVM achieves stable and higher classification accuracy. 
This is mainly due to the enhanced suitability of the hyper-planes logic in identifying and constraining the clusters across the different MFCC components, while getting rid of unstable readings. 
Despite the excellent achieved results, we remark that this paper aims to provide experimental evidence that the sound emitted by the drone can be tightly correlated with the payload weight the drone is carrying, and not to find the most suitable classifier to use for this problem---this objective is left for future work.

\begin{figure*}
     \centering
     \begin{subfigure}[b]{0.49\textwidth}
         \centering
         \includegraphics[width=\columnwidth]{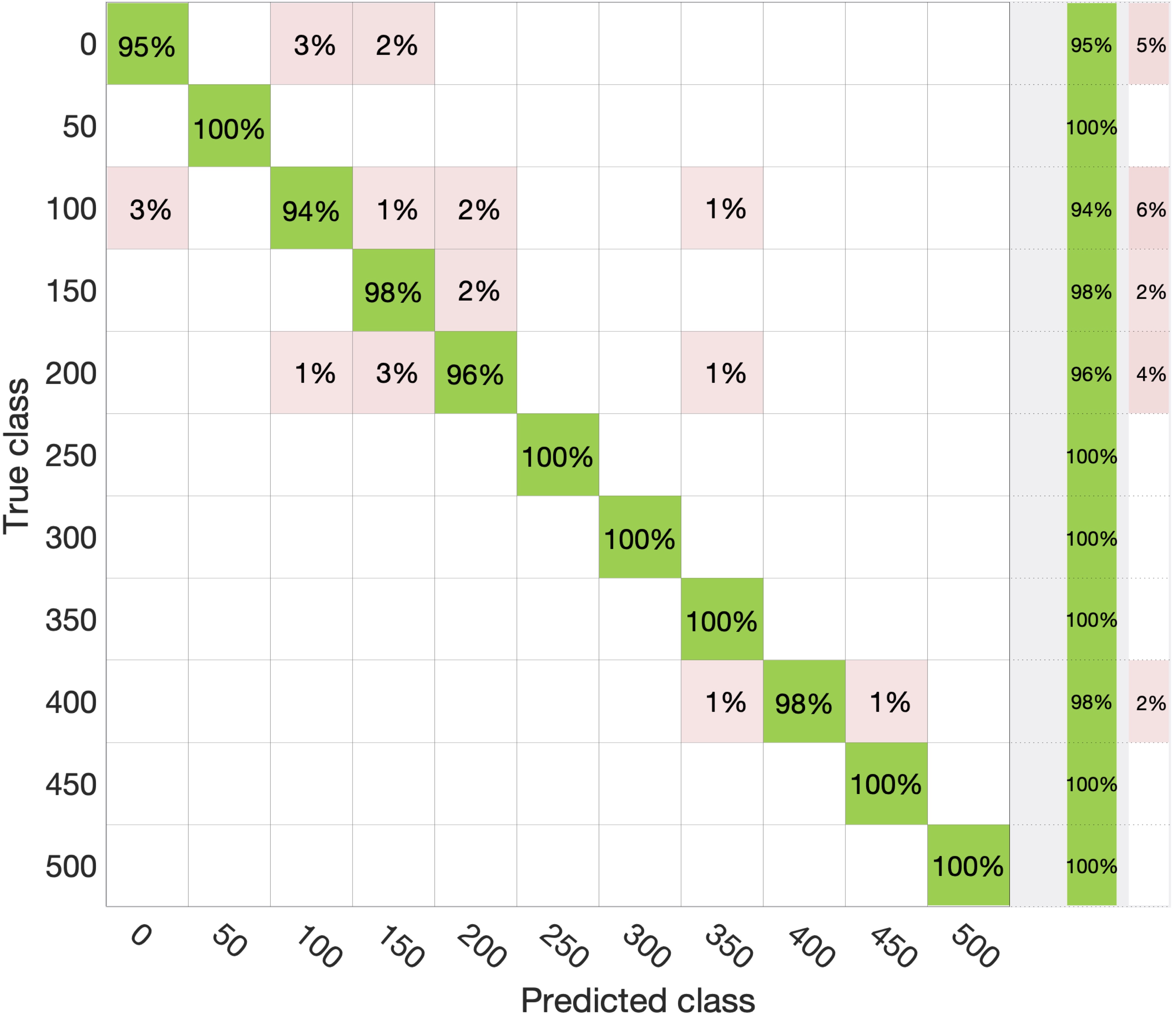}
         \caption{Confusion matrix obtained using a time window $t=0.25$~s for the computation of the MFCC components.}
         \label{fig:CM0.25s}
     \end{subfigure}
     \hfill
     \begin{subfigure}[b]{0.49\textwidth}
         \centering
         \includegraphics[width=\textwidth]{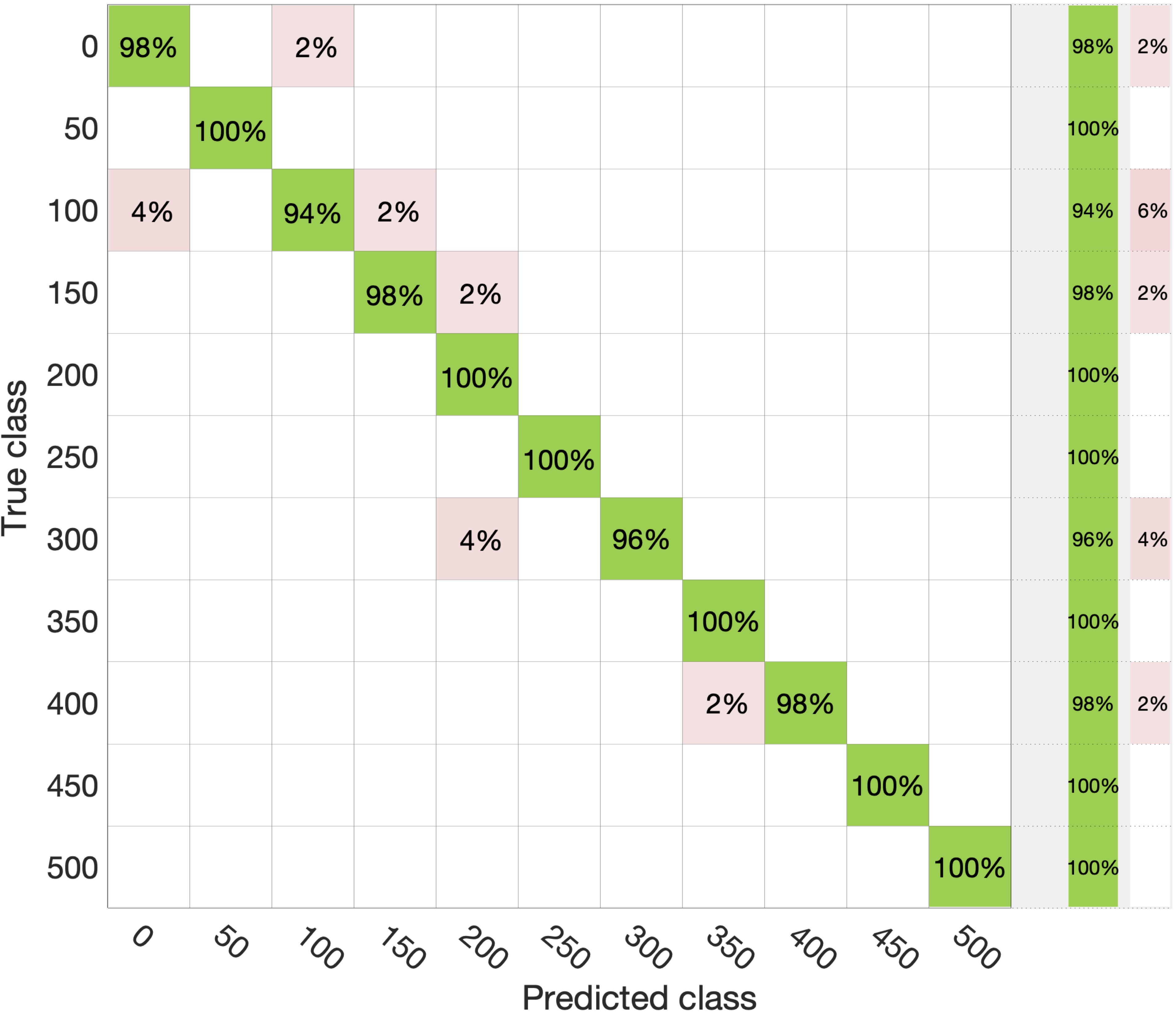}
         \caption{Confusion matrix obtained using a time window $t=1$~s for the computation of the MFCC components.}
         \label{fig:CM1s}
     \end{subfigure}
     \caption{Confusion matrices resulting from the application of the cubic SVM classifier to the identification of the weight of the payload carried by the drone using its acoustic noise.}
     \label{fig:CM}
     
\end{figure*}

Finally, for the sake of completeness and to provide further insights, in Figure~\ref{fig:CM}, we show the confusion matrices resulted from the application of the cubic SVM classification tool when considering the MFCC components extracted over recordings of $0.25$~s and $1$~s, respectively.

Looking at the reported results, we notice that most of the few samples that are wrongly classified are obtained among recordings with adjacent payload weight classes. This is a reasonable and expected finding, as the motors and blades rotation speeds are very similar when the drone is carrying similar payload weights. Looking at Figure~\ref{fig:CM1s}, we also notice that all the errors related to non-adjacent classes are removed when the recording time increases to $1$~s, further highlighting the beneficial effect of increased recording time in getting rid of short and transient noise fluctuation effects.


\textbf{Noise Analysis.} As previously discussed in Section~\ref{sec:setup}, the sound recordings of the drone carrying different payload weights have been collected in a standard outdoor environment, characterized by noise typical of an urban scenario, such as birds sounds, cars passing in the background, wind, and people voices, to name a few. 
However, we could mostly get rid of these noise components by pointing the directional microphone toward the drone, 
in a way that the sound produced by the drone would be dominant. 
However, in real use-cases as the ones introduced in Section~\ref{sec:scenario}, several factors could reduce the SNR of the recordings. For instance, the orientation of the microphones could not be optimal, or the distance of the drone from the recording equipment could not be the most suitable one, leading to an increased noise level.

\begin{figure}[htbp!]
\centerline{\includegraphics[width=\linewidth]{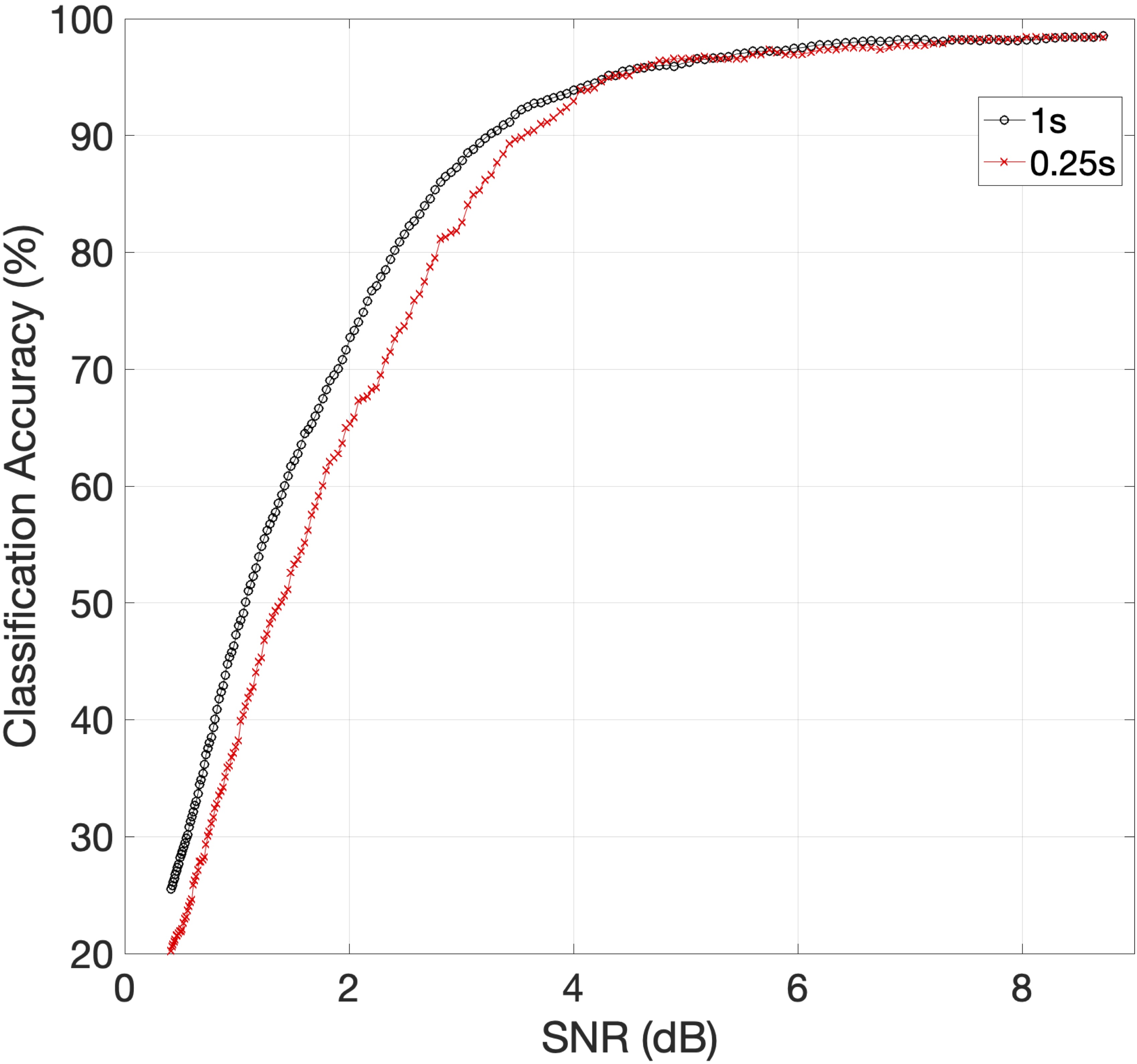}}
\caption{Classification accuracy in the detection of the payload weight carried by the drone, with different \ac{SNR} values and different recording times. }
\label{fig:SNR}
\end{figure}

To take into account the effect of the noise on the classification accuracy reported by our methodology, we studied the effect of an increasing level of noise on the reported classification accuracy. Considering as a reference the cubic-SVM classification tool and the two time-windows previously considered for the sound recordings, we perturbed each of the traces corresponding to the sound of the drone with noise following the \ac{AWGN} model, consistently with other scientific contributions working on the sound emitted by the drone~\cite{shi2018hidden},~\cite{varghees2014_bspc}.
To carry out these tests, we considered a total number of $183$ \ac{SNR} levels, resulting in \ac{SNR} values from $0$~dB to $8.8$~dB. Figure~\ref{fig:SNR} reports the classification accuracy achieved by the cubic-SVM tool for the different tested SNR levels, by considering the time-windows of $0.25$~s and $1$~s.

We can notice that, as expected, the classification accuracy increases with the increase in the SNR, reporting an exponential behavior. Considering the worst-case time-window of $0.25$~s for the MFCC computation, we highlight that the accuracy of the cubic-SVM classification tool reaches the value of $98.4$\% obtained in our previous analysis for a minimum SNR value of $8$~dB. This means that at least $8$~dB of SNR is required for the sound recording to obtain the remarkable classification performances previously discussed. Depending on the required classification accuracy, this lower bound could be further decreased (for instance, the $95$\% classification accuracy is reached starting from an SNR of $4.5$~dB). Finally, we notice that increasing the time-window used for the recordings and the computation of the MFCC components have a beneficial effect in the robustness to the noise, with slightly higher performances (approximately $5$\% in the range $\left[1 , 4.5 \right]$~dB ).


\section{Conclusion}
\label{sec:conc}

In this paper, to the best of our knowledge, we have been the first to show the tight relationship existing between the payload weight carried by a drone and its unique acoustic fingerprint. 
We experimentally proved that the variation of the thrust required for a drone to hover when carrying different payload weights leads to differences in the profile of the sound emitted by the drone, that can be used to obtain an estimation of the weight that the specific drone is carrying. 
Our analysis, conducted using recordings of approximately $170$~seconds for eleven different weights carried by a 3DR SOLO drone, from 0~g to 500~g, shows significant differences in the pitch of the produced sound, caused by the different speed of the motors with different payload weights. 
In a constrained application scenario, we demonstrated that the pitch of the sound could be used to identify if a drone is carrying a payload, or not. 
In particular, recording the sound for $2.5$~s, we showed that a negligible error rate can be achieved if the drone carries $150$~g of payload or more. When the analysis is carried out on a normal laptop (relaxing the computational constraints), MFCC components can be used in combination with a supervised ML classifier, to obtain remarkable classification accuracy in the identification of the specific payload weight the drone is carrying. In this context, we showed that a cubic SVM classifier can achieve $98.4$\% classification accuracy over a recording time of only $0.25$~s, while the performances further improve when the recording time can be increased. Additional analysis aimed at assessing the impact of noise over the achieved results showed that such a remarkable classification accuracy can be obtained when the sound of the drone is characterized by a minimum \acl{SNR} of $8$~dB.

All the data used for our analysis have been released as open-source, to enable practitioners, Industry, and Academia, to independently validate our findings and to use our data as a ready-to-use basis for further investigations.


\section*{Acknowledgments}
This publication was partially supported by awards NPRP-S-11-0109-180242, NPRP X-063-1-014, and GSRA6-1-0528-19046, from the QNRF-Qatar National Research Fund, a member of The Qatar Foundation. The information and views set out in this publication are those of the authors and do not necessarily reflect the official opinion of the QNRF.

\balance
\bibliographystyle{IEEEtran}
\bibliography{references}
\balance

\begin{IEEEbiography}[{\includegraphics[width=1in,height=1.15in,clip,keepaspectratio]{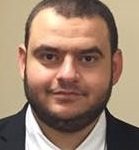}}]{Omar Adel Ibrahim}
Omar Ibrahim is currently a Computer Science and Engineering Ph.D. candidate at HBKU-CSE-ICT, Doha, Qatar. He received his bachelor’s degree in computer engineering from Qatar University in 2017 and the Master of Cybersecurity degree from HBKU-CSE-ICT in 2019. His main research interests cover security issues in Cyber-Physical Systems (CPS), including Drones, GPS, and USB devices security.
\end{IEEEbiography}

\begin{IEEEbiography}[{\includegraphics[width=1in,height=1.15in,clip,keepaspectratio]{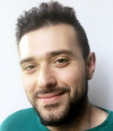}}]{Savio Sciancalepore}
Savio Sciancalepore is currently Post Doc at HBKU-CSE-ICT, Doha, Qatar. He received his master degree in Telecommunications Engineering in 2013 and the PhD in 2017 in Electric and Information Engineering, both from the Politecnico di Bari, Italy. He received the prestigious award from the ERCIM Security, Trust, and Management (STM) Working Group for the best Ph.D. Thesis in Information and Network Security in 2018. His major research interests cover security issues in Cyber-Physical Systems (CPS), including Internet of Things (IoT), Avionic Communication Technologies, Drones Communications, Intrusion Detection, and Network Security.
\end{IEEEbiography}

\begin{IEEEbiography}[{\includegraphics[width=1in,height=1.25in,clip,keepaspectratio]{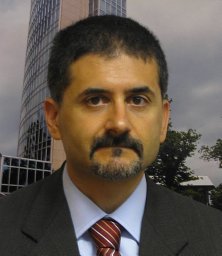}}]{\textbf{Roberto Di Pietro}} is Full Professor in Cybersecurity at HBKU-CSE. Previously, he was in the capacity of Global Head Security Research at Nokia Bell Labs, and Associate Professor of Computer Science at University of Padova, Italy. He has been working in the security field for more than 20 years, leading both technology-oriented and research-focused teams in the private sector, government, and academia (MoD, United Nations HQ, EUROJUST, IAEA, WIPO). His main research interests include security and privacy for wired and wireless distributed systems (e.g. Blockchain technology, Cloud, IoT, WSN, RFID, OSNs), virtualization security, applied cryptography, computer forensics, and data science. He is serving as an AE for Elsevier ComCom and other Intl. journals. In 2011-2012 he was awarded a Chair of Excellence from University Carlos III, Madrid. He has been publishing 190+ scientific papers over these topics, co-authored two books, and contributed to a few others. Google says that he has been receiving 7500+ citations, with an h-index=42 and an i-index=116.
\end{IEEEbiography}

\end{document}